\documentclass [10pt,twocolumn]{IEEEtran}
\usepackage{algorithm,algorithmic,amsbsy,amsmath,amssymb,epsfig,bbm,mathrsfs,multirow,amsthm}
\usepackage[T1]{fontenc}
\usepackage[utf8]{inputenc}
\usepackage{amsmath}
\usepackage{algorithm}
\usepackage{array,multirow,graphicx}
\usepackage{color}
\usepackage{cite}
\usepackage{float}
\usepackage{makecell}
\usepackage[x11names,dvipsnames,table]{xcolor} 
\usepackage{colortbl}
\usepackage{multirow}
\usepackage{lipsum}
\usepackage{graphicx} 
\usepackage{booktabs}
\usepackage{tabularx} 
\usepackage[colorlinks=true,linkcolor=black,citecolor=colorlinks=true,linkcolor=black,citecolor=blue,urlcolor=blue,urlcolor=blue]{hyperref}

\definecolor{mygray}{gray}{0.6}
\definecolor{myblue}{rgb}{0.8,0.85,1} 
\usepackage{array}

\newcolumntype{L}[1]{>{\raggedright\let\newline\\\arraybackslash\hspace{0pt}}m{#1}}
\newcolumntype{C}[1]{>{\centering\let\newline\\\arraybackslash\hspace{0pt}}m{#1}}
\newcolumntype{R}[1]{>{\raggedleft\let\newline\\\arraybackslash\hspace{0pt}}m{#1}}

\usepackage{array, tabularx, boldline}
\usepackage{eurosym}
\usepackage{amstext} 
\DeclareRobustCommand{\officialeuro}{%
  \ifmmode\expandafter\text\fi
  {\fontencoding{U}\fontfamily{eurosym}\selectfont e}}

\usepackage{array, tabularx, boldline}
\usepackage{graphicx}
\usepackage{cellspace}
\begin{document}
\title{\huge From Denoising to Decision Making: A Survey on Diffusion Model-Enabled Deep Reinforcement Learning for Wireless Networks}

\author{Nguyen Cong Luong,~\IEEEmembership{Member,~IEEE}, Zeping Sui,~\IEEEmembership{Member,~IEEE}, Jie Cao,~\IEEEmembership{Member,~IEEE}, Min Xu, Nguyen Duc Hai, Zhihao Dong, Nguyen Duc Duy Anh, Qiushi Zhao, Nguyen Quoc Khanh, Zhe Fu, Shaohan Feng,~\IEEEmembership{Member,~IEEE}, and Bo Ma,~\IEEEmembership{Member,~IEEE}

\thanks{Nguyen Cong Luong, Nguyen Duc Duy Anh, and Nguyen Quoc Khanh are with the Phenikaa School of Computing, Phenikaa University, Hanoi 12116, Vietnam. (e-mail: \{luong.nguyencong, 21011488, khanh.nguyenquoc\}@phenikaa-uni.edu.vn).}

\thanks{Zeping Sui is with the School of Computer Science and Electronics Engineering, University of Essex, Colchester CO4 3SQ, U.K. (e-mail: zepingsui@outlook.com).}

\thanks{Jie Cao and Zhihao Dong are with the School of Information Science and Technology, Harbin Institute of Technology, Shenzhen 518055, China. (e-mail: \{caojhitsz, zhihaodong\}@ieee.org).}

\thanks{Min Xu is with the School of Mathematics, Statistics and Mechanics, Beijing University of Technology, Beijing 100124, China. (e-mail: xm@bjut.edu.cn).}

\thanks{Nguyen Duc Hai is with the Faculty of Artificial Intelligence and Data Science, Phenikaa University, Duong Noi, Hanoi 12116, Vietnam. (e-mail: hai.nguyenduc@phenikaa-uni.edu.vn).}

\thanks{Qiushi Zhao, Shaohan Feng, and Bo Ma are with the School of Information and Electronic Engineering (Sussex Artificial Intelligence Institute), Zhejiang Gongshang University, Hangzhou 310018, China. (e-mail: \{25020090093, feng\_shaohan, mabo\}@mail.zjgsu.edu.cn).}

\thanks{Zhe Fu is with the School of Integrated Circuits, Guangdong University of Technology, Guangzhou, 510000, Guangdong, China. (e-mail: fuzhe@gdut.edu.cn).}


}

\maketitle
\begin{abstract}
Deep reinforcement learning (DRL) has long been a promising solution for sequential resource management in wireless networks. However, conventional DRL methods are fundamentally limited by their reliance on unimodal policy distributions, inefficient exploration in high-dimensional action spaces, and poor adaptability to dynamic and heterogeneous environments. Meanwhile, diffusion models (DMs) as one of the most powerful families of generative AI have demonstrted remarkable capabilities in modeling complex, multi-modal data distributions across diverse domains. The integration of DMs and DRL has opened a new and rapidly growing research direction, in which DM-enabled policies substantially enhance decision quality by capturing the complex, discontinuous, and multimodal action structures inherent in wireless resource management. In this paper, we present a comprehensive survey of DM-enabled DRL algorithms and their applications for various issues in wireless networks. Particularly, we first provide the theoretical background of DM and present different DM-enabled DRL algorithms. We then systematically review applications of DM-enabled DRL for across computation offloading in mobile edge computing, UAV-assisted, vehicular, and AIGC-driven systems, as well as wireless resource allocation, physical-layer security, and robotics and UAV planning. We conclude the paper by higlight future research directions.

\end{abstract}


\section{Introduction}

The rapid growth of artificial intelligence (AI) services is reshaping the landscape of next-generation wireless, edge, and cloud computing systems. Emerging applications such as AI-generated content (AIGC) \cite{ye2024optimizing}, immersive metaverse services \cite{wang2022survey}, Large Language Model (LLM) inference \cite{zhang2024edgeshard}, autonomous driving \cite{liu2022mobility}, and uncrewed aerial vehicle (UAV)-assisted systems \cite{huang2023unmanned} require not only high communication rates, but also low-latency computation, adaptive resource orchestration, and reliable decision making under dynamic environments. Therefore, optimization problems in wireless networks, especially AI-assisted ones, are becoming more complex and challenging. 

Traditional optimization-based approaches have achieved notable success in wireless network optimization. However, they often rely on accurate system models, convexity assumptions, or tractable problem structures. In practice, however, wireless networks are characterized by time-varying channels, stochastic task arrivals, heterogeneous service demands, mixed discrete-continuous decision variables, and non-convex objectives. Thus, deep reinforcement learning (DRL) has emerged as a promising tool for sequential resource management, as it can learn adaptive policies directly from interaction data without requiring complete analytical models. Representative algorithms such as Deep-Q Networks (DQN) \cite{mnih2013playing}, Deep Deterministic Policy Gradient (DDPG) \cite{lillicrap2015continuous}, Soft-Actor Critic (SAC) \cite{haarnoja2018soft}, Proximal Policy Optimization (PPO) \cite{schulman2017proximal}, and Multi-Agent DDPG (MADDPG) \cite{lowe2017multi}, have been widely adopted for computation offloading, power control, spectrum allocation, routing, and multi-agent coordination. Nevertheless, conventional DRL methods usually parameterize policies with simple distributions, such as Gaussian actors or categorical distributions, which may be insufficient for representing the multi-modal and high-dimensional action structures commonly observed in AI service networks.


To address the weaknesses of conventional DRL methods, difusion models (DMs) \cite{sohl2015deep, ho2020denoising} have recently emerged as a promising solution. DMs, which have enjoyed a wide range of applications across various domains, including computer vision \cite{croitoru2023diffusion}, natural language processing \cite{zou2023survey}, and multi-modal modeling \cite{jiang2024survey}, 
offer several notable advantages for DRL integration. First, they can capture complex and multimodal action distributions, which conventional Gaussian or deterministic policy networks often fail to represent, enabling more expressive policy parameterization. Second, DMs allow diverse action sampling due to their stochastic denoising process, which can improve exploration in high-dimensional and non-convex environments. Third, DMs can generate entire action sequences and trajectories, making them suitable for long-horizon planning and sequential decision making. Fourth, DMs are highly compatible with offline DRL due to their ability to keep generated actions close to the behavior data manifold, thus reducing out-of-distribution actions and mitigating distribution shift. Finally, DM-enabled policies are flexible enough to incorporate complex system constraints through guided sampling, projection, or feasibility-aware denoising. 

The integration of DMs and DRL has opened a new direction for optimization in wireless systems, in which DMs can serve multiple roles. First, DM-enabled DRL algorithms can be effective in computation offloading and edge-cloud scheduling, due to their ability to learn multi-modal and nonlinear offloading and service-migration decisions, which is infeasible for conventional Gaussian policies \cite{du2024integrated, rao2025computation, wang2024dmais}. Second, DM-enabled DRL can improve AIGC service orchestration, since diffusion policies can capture heterogeneous service demands, dynamic user mobility, as well as hybrid discrete-continuous decisions such as model selection, cache/cloud routing, and DNN partitioning \cite{du2024diffusion, yao2025enhancing, yang2025diffusion}. Third, DM-enabled DRL is a feasible solution for wireless resource allocation by generating diverse candidate actions and overcoming the unimodal action limit of conventional DRL schemes under conditions such as dynamic channels and imperfect CSI \cite{zhang2025improve, wang2025uplink, khoramnejad2025carrier}. Fourth, DM-enabled DRL can enhance security, privacy, and robustness: the DMs can act either as a robust action generator under adversarial interference and jamming, or as a privacy-preserving state transformation module \cite{zhang2024multi, liang2024uav}. Fifth, DM-enabled DRL can benefit routing and sensing schemes by serving as a trajectory-level planner that generates temporally coherent action sequences, movement policies, or high-level goals with respect to mobility, energy, and long-horizon coordination constraints \cite{shi2025diffusion, qiao2025combined}.

There have been some existing surveys regarding DM-enabled DRL algorithms, and DRL algorithms for wireless systems. However, they do not provide a comprehensive literature review on the integration of DM-enabled DRL algorithms in wireless networks. The authors in \cite{du2024enhancing} provided a basic background on DMs, followed by a series of case studies of DM-enabled techniques, including DRL, incentive mechanisms, semantic communications (SemCom) and Internet-of-Vehicles (IoV) networks. In contrast, the works in \cite{luong2025diffusion, fan2026generative} broadly reviewed DM-enabled techniques for wireless networks and future communication systems. The authors in \cite{zhu2023diffusion} surveyed DM-enabled DRL approaches and classified the roles of DMs in RL, such as policy modeling, planning, and data synthesis. However, it is not focused on wireless networks, edge computing, or resource management. The work in \cite{letafati2023diffusion} discussed DMs for wireless systems, with a focus on channel modeling, denoising, reconstruction, and SemCom. Meanwhile, the authors in \cite{luong2019applications} reviewed the applications of DRL in emerging issues in communications and networking. This line of literature has been extended by \cite{bai2023toward} and \cite{zhao2025survey}, in which DRL algorithms for UAV networks were extensively discussed, and \cite{alwarafy2021deep}, which focused on DRL-assisted radio resource allocation and management. More recently, the authors in \cite{hazra2024drl} summarized the importance of DRL methods for wireless edge networks, as well as DRL-enabled IoT applications for edge systems. However, none of the prior works specifically provided a comprehensive review of DM-enabled DRL for wireless networks.




The main contributions of this survey are summarized as follows.
\begin{itemize}
    \item We provide a structured overview of the foundations of DM-enabled DRL, including denoising diffusion probabilistic models (DDPMs), score-based generative models (SGMs), guided sampling, diffusion policies, DQN, diffusion actor-critic methods, diffusion PPO, DM-enabled multi-agent RL, trajectory-level diffusion planning, data augmentation, reward shaping, hierarchical learning, and transfer learning.
    \item We present a comprehensive taxonomy of DM-enabled DRL methods that have recently used for wireless networks. The reviewed methods are categorized according to their algorithmic roles, including policy generation, action refinement, trajectory planning, replay-buffer augmentation, reward shaping, privacy-preserving state transformation, and domain-adaptive transfer.
    \item We systematically review the applications of DM-enabled DRL in computation offloading and resource management, covering mobile edge computing, UAV-assisted offloading, vehicular edge computing, AIGC and LLM service offloading, edge-cloud scheduling, and network design. 
    \item We further investigate DM-enabled DRL for wireless resource allocation, security, privacy, routing, robotics control, vehicle routing, and UAV planning. 
    \item We discuss future research directions, including quantization-aware diffusion policies for resource-constrained wireless networks, lightweight and real-time diffusion inference, safety-constrained DM-enabled DRL for critical infrastructure, multi-agent DM-enabled DRL with scalable coordination, risk-sensitive and freshness-aware DM-enabled DRL, federated and privacy-preserving DM-enabled DRL, and foundation model and digital twin-assisted DM-enabled DRL.
\end{itemize}
The remainder of this paper is organized as follows. Section \ref{sec:tutorial} introduces the background and methods of DM-enabled DRL. Section \ref{sec:computation_offloading} reviews DM-enabled DRL for computation offloading. Section \ref{sec:resource_security_routing} discusses applications of DM-based DRL in wireless resource allocation, security, and routing issues. Finally, Section \ref{sec:conclusions} concludes the paper, summarizing the major lessons learned and discussing future research directions. A summarization of the main abbreviations can be found in Table \ref{tab:main_abbreviations}.

\begin{table}[t]
\centering
\caption{Main abbreviations.}
\label{tab:main_abbreviations}
\scriptsize
\begin{tabular}{|l|l|}
\hline
\textbf{Abbreviation} & \textbf{Definition} \\
\hline
AI & Artificial Intelligence \\
\hline
AIGC & AI-Generated Content \\
\hline
AoI & Age of Information \\
\hline
CSI & Channel State Information \\
\hline
DDPM & Denoising Diffusion Probabilistic Model \\
\hline
DDPG & Deep Deterministic Policy Gradient \\
\hline
DM-enabled DRL & DM-enabled Deep Reinforcement Learning \\
\hline
DM & Diffusion Model \\
\hline
DNN & Deep Neural Network \\
\hline
DQN & Deep Q-Network \\
\hline
DRL & Deep Reinforcement Learning \\
\hline
GAI & Generative Artificial Intelligence \\
\hline
GDM & Generative Diffusion Model \\
\hline
IoT & Internet of Things \\
\hline
IRS & Intelligent Reflecting Surface \\
\hline
ISAC & Integrated Sensing and Communication \\
\hline
LEO & Low Earth Orbit \\
\hline
LLM & Large Language Model \\
\hline
MADDPG & Multi-Agent Deep Deterministic Policy Gradient \\
\hline
MADRL & Multi-Agent Deep Reinforcement Learning \\
\hline
MARL & Multi-Agent Reinforcement Learning \\
\hline
MEC & Mobile Edge Computing \\
\hline
MDP & Markov Decision Process \\
\hline
PPO & Proximal Policy Optimization \\
\hline
PRB & Physical Resource Block \\
\hline
QoE & Quality of Experience \\
\hline
QoS & Quality of Service \\
\hline
RL & Reinforcement Learning \\
\hline
RSMA & Rate-Splitting Multiple Access \\
\hline
SAC & Soft Actor-Critic \\
\hline
TD3 & Twin Delayed Deep Deterministic Policy Gradient \\
\hline
UAV & Unmanned Aerial Vehicle \\
\hline
UGV & Unmanned Ground Vehicle \\
\hline
UE & User Equipment \\
\hline
V2I & Vehicle-to-Infrastructure \\
\hline
V2V & Vehicle-to-Vehicle \\
\hline
VEC & Vehicular Edge Computing \\
\hline
VR & Virtual Reality \\
\hline
\end{tabular}
\vspace{-2em}
\end{table}

\section{Background and Key Methods}
\label{sec:tutorial}
\subsection{Foundations of Diffusion Models} 
DMs are a powerful class of generative models that learn to gradually denoise data, thereby producing new samples from pure noise. Meanwhile, DRL is a paradigm for optimizing policies in Markov Decision Processes (MDPs) by maximizing expected cumulative rewards. The motivation for integration is that DMs can capture complex, multi-modal distributions with high stability, while DRL allows optimizing objectives beyond likelihood, such as action quality, diversity, or reward alignment. Combining the DM and DRL, i.e., DM-
enabled DRL, leverages two strengths: diffusion policies can imitate expert behavior or generate action sequences with high returns, while DRL can fine-tune diffusion models to satisfy specific reward signals. We next provide the theoretical foundations of DMs, including the two predominant types and the guided sampling techniques relevant to DRL tasks.

\begin{figure*}[htbp]
    \centering
    \includegraphics[width=\textwidth]{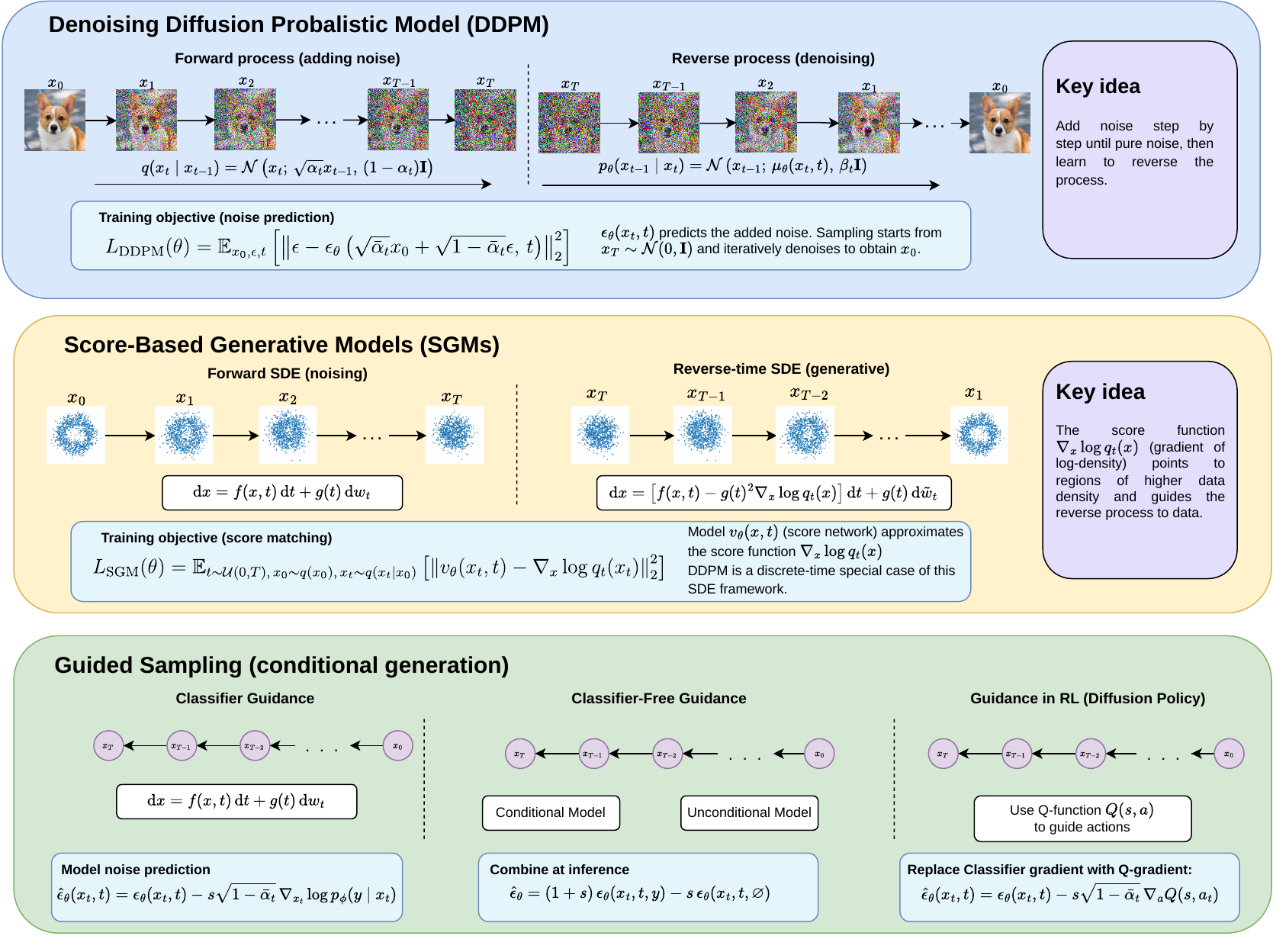}
    \caption{Theoretical foundations of diffusion models and conditional generation mechanisms. The diagram illustrates the discrete-time forward and reverse Markov chains in DDPMs (top), the continuous-time stochastic differential equations underpinning SGMs (middle), and the guided sampling paradigms, highlighting the integration of Q-function gradients to steer action generation in DM-enabled RL policies (bottom).}
    \label{fig:foundationDM}
\end{figure*}

\subsubsection{Denoising Diffusion Probabilistic Model}
A denoising diffusion probabilistic model (DDPM) \cite{ho2020denoising} makes use of a parameterized diffusion process consisting of two Markov chains for data corruption and restoration, respectively. Given real data $x_0 \sim q(x_0)$, the forward chain generates $T$ random variables $x_1, x_2, \ldots, x_T$ by gradually injecting independent Gaussian noise. Specifically, at each diffusion step $t$, let $\epsilon_t \sim \mathcal{N}(\mathbf{0},\mathbf{I})$ denote an independent standard Gaussian noise variable. The forward transition is then given by
\begin{equation}
    \label{noise_addition}
    x_t = \sqrt{\alpha_t}x_{t-1} + \sqrt{1-\alpha_t}\,\epsilon_t, \quad \epsilon_t \sim \mathcal{N}(\mathbf{0},\mathbf{I}),
\end{equation}
with $\alpha_t = 1 - \beta_t$ and noise schedule $\{\beta_t\}_{t=1}^T$. Denoting $\bar{\alpha}_t = \prod_{i=1}^{t}\alpha_i$, one can directly sample any noisy step from the clean data by using another independent standard noise variable $\epsilon \sim \mathcal{N}(\mathbf{0},\mathbf{I})$:
\begin{equation}
    \label{mapping}
    x_t = \sqrt{\bar{\alpha}_t}\,x_0 + \sqrt{1-\bar{\alpha}_t}\,\epsilon, \quad \epsilon\sim\mathcal{N}(\mathbf{0},\mathbf{I}).
\end{equation}
As $\bar{\alpha}_T\to 0$, the marginal $q(x_T)\approx\mathcal{N}(\mathbf{0},\mathbf{I})$. The reverse Markov chain recovers $x_0$ from noise via the learnable transition kernel
\begin{equation}
    \label{reverse_kernel}
    p_\theta(x_{t-1}|x_t) = \mathcal{N}\!\left(\frac{1}{\sqrt{\alpha_t}}\!\left(x_t - \frac{\beta_t}{\sqrt{1-\bar{\alpha}_t}}\epsilon_\theta(x_t,t)\right),\,\beta_t\mathbf{I}\right),
\end{equation}
where $\epsilon_\theta$ is trained by minimizing the mean-squared denoising error:
\begin{equation}
    \label{loss_elbo}
    L_{\mathrm{DDPM}}(\theta) = \mathbb{E}_{x_0,\epsilon,t}\!\left[\left\|\epsilon - \epsilon_\theta\!\left(\sqrt{\bar{\alpha}_t}x_0+\sqrt{1-\bar{\alpha}_t}\epsilon,\,t\right)\right\|^2\right].
\end{equation}

\subsubsection{Score-Based Generative Models}
Score-based generative models (SGMs) \cite{song2020score} generalize DDPMs by describing the forward noising process as a stochastic differential equation (SDE):
\begin{equation}
    \label{sde}
    \mathrm{d}x = f(x,t)\,\mathrm{d}t + g(t)\,\mathrm{d}w,
\end{equation}
where $f(x,t)$ is a pre-specified drift, $g(t)$ is the diffusion coefficient, and $w$ is the standard Wiener process. The corresponding reverse SDE that recovers clean data is
\begin{equation}
    \label{rev_sde}
    \mathrm{d}x = \bigl[f(x,t)-g(t)^2\,\nabla_x\log q_t(x)\bigr]\,\mathrm{d}t + g(t)\,\mathrm{d}\tilde{w},
\end{equation}
where $\tilde{w}$ is the reverse-time Wiener process and $\nabla_x\log q_t(x)$ is the \textit{score function} of the marginal density $q_t$. A score network $v_\theta$ approximates this function by minimizing
\begin{equation}
    \label{sgm_loss}
    L_{\mathrm{SGM}}(\theta) = \mathbb{E}_{x_0,t,x_t}\!\left[\left\|v_\theta(x_t,t)-\nabla_{x_t}\log p(x_t|x_0)\right\|_2^2\right].
\end{equation}

\subsubsection{Guided Sampling}
Guided sampling produces samples from a conditional distribution $p(x|y)$, where $y$ encodes desired properties of the generated output, such as a target reward value or a conditioning state. There are two main approaches. In \textit{classifier guidance}, a differentiable classifier $p_\phi(y|x)$ is incorporated into the reverse process, yielding the modified noise prediction
\begin{equation}
    \hat{\epsilon}_\theta(x_t,t) = \epsilon_\theta(x_t,t) - s\,\sqrt{1-\bar{\alpha}_t}\,\nabla_{x_t}\log p_\phi(y|x_t),
\end{equation}
where $s>0$ is the guidance scale. \textit{Classifier-free guidance} \cite{ho2022classifier} jointly trains a conditional model $\epsilon_\theta(x_t,t,y)$ and an unconditional model $\epsilon_\theta(x_t,t,\varnothing)$, where $\varnothing$ indicates the absence of the state conditioning corresponding to a unconditional policy. Then, $\epsilon_\theta(x_t,t y)$ and $\epsilon_\theta(x_t, t, \varnothing)$ are combined at inference as $\hat{\epsilon}_\theta = (1+s)\epsilon_\theta(x_t,t,y) - s\,\epsilon_\theta(x_t,t,\varnothing)$, where $y$ is the conditioning signal. In DM-enabled DRL, $y$ is typically the agent's current state $s$, and thus the policy generates actions $a$ conditioned on $s$. When a Q-function is available, its gradient replaces the classifier gradient to steer denoising toward high-reward actions, as detailed in the following methods.

\subsection{DM-enabled DRL}

The general strategy for DM-enabled DRL is to replace the conventional stochastic policy $\pi_\theta(a|s)$, typically parameterized as a unimodal Gaussian, with a DDPM-based policy. Given state $s$, an action $a_0$ is produced by first sampling $a_T\sim\mathcal{N}(\mathbf{0},\mathbf{I})$ and then iteratively applying the learned reverse kernel:
\begin{equation}
    \label{eq:diffusion_reverse}
    \begin{aligned}
        a_{k-1} = \frac{1}{\sqrt{\alpha_k}}\!\left(a_k - \frac{\beta_k}{\sqrt{1-\bar{\alpha}_k}}\epsilon_\theta(a_k, k, s)\right) + \sqrt{\beta_k}\,\xi,\\ \xi\sim\mathcal{N}(\mathbf{0},\mathbf{I}),
    \end{aligned}
\end{equation}
for $k = T, T-1, \ldots, 1$, where $\epsilon_\theta(a_k, k, s)$ is a noise-prediction network conditioned on both the current noisy action $a_k$ and the state $s$. This parameterization endows the policy with multi-modal expressiveness: a single state can map to a distribution over diverse high-quality actions, which is critical for the complex resource management and control problems studied in this survey.
\begin{figure}[htbp]
    \centering
    \includegraphics[width=\linewidth]{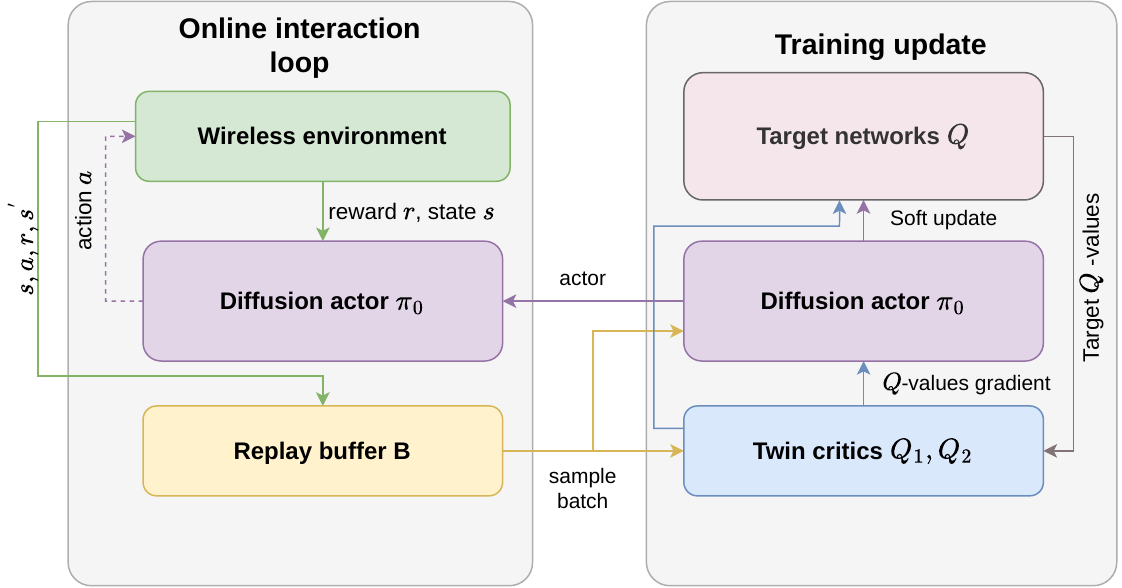}
    \caption{The overall architecture and training workflow of the proposed off-policy DM-enabled DRL framework, comprising the online interaction loop with the wireless environment and the offline training update phase featuring actor-critic optimization and soft updates for target networks.}
    \label{fig:DMDRL}
\end{figure}
We now describe the DM-enabled DRL methods that appear most frequently in the literature reviewed in Sections~\ref{sec:computation_offloading} and~\ref{sec:resource_security_routing}.

\subsubsection{DM-enabled Q-Learning}

DM-enabled Q-Learning (DM-enabled-QL) \cite{wang2022diffusion} is the foundational offline DM-enabled DRL algorithm and is widely adopted across the surveyed papers. The policy is a DDPM parameterized by $\theta$, trained on a static offline dataset $\mathcal{D}$ by minimizing a composite objective that simultaneously enforces behavioral regularization and policy improvement:
\begin{equation}
    \label{diffusion_ql}
    L(\theta) = L_{\mathrm{DDPM}}(\theta) - \lambda\,\mathbb{E}_{s\sim\mathcal{D},\,a_0\sim\pi_\theta(\cdot|s)}\!\left[Q_\phi(s, a_0)\right],
\end{equation}
where $L_{\mathrm{DDPM}}$ is the standard DDPM denoising loss in \eqref{loss_elbo}, $Q_\phi$ is a twin critic trained via Bellman updates on $\mathcal{D}$, and $\lambda>0$ balances the two objectives. The first term ensures that the learned policy is close to the behavior policy that generated $\mathcal{D}$, preventing out-of-distribution action extrapolation - a critical concern in offline DRL. The second term pushes the policy toward regions of high Q-value. At inference, Q-function guidance is incorporated into each denoising step as
\begin{equation}
    \hat{\epsilon}_\theta(a_k, k, s) = \epsilon_\theta(a_k, k, s) - \eta_k\,\nabla_{a_k} Q_\phi(s, a_k),
\end{equation}
where $\eta_k$ is a step-size schedule. This formulation makes DM-enabled QL particularly attractive for offline resource management and spectrum allocation tasks, where safe exploitation of pre-collected operational data is essential.

\subsubsection{DM-enabled Soft Actor-Critic}

Soft Actor-Critic (SAC) \cite{haarnoja2018soft} maximizes the entropy-regularized objective
\begin{equation} \label{eq:entropy_dsac}
    J(\pi) = \mathbb{E}_{s\sim\mathcal{D},\,a\sim\pi}\!\bigl[Q(s,a) - \alpha\log\pi(a|s)\bigr],
\end{equation}
where $\alpha$ is a temperature parameter that encourages exploration. DM-enabled SAC (DSAC) replaces the Gaussian actor with a DDPM policy \cite{wang2024diffusion}. Since the closed-form log-likelihood $\log\pi_\theta(a|s)$ is intractable for a diffusion policy, the entropy term is approximated using the DDPM loss as a tractable proxy. The actor update optimizes
\begin{equation}
    \label{dsac_actor}
    L_{\mathrm{actor}}(\theta) = -\mathbb{E}\!\left[\min(Q_1(s,a_0),\, Q_2(s,a_0))\right] + \mathcal{L}_{\mathrm{DDPM}}(\theta),
\end{equation}
where twin critics $Q_1, Q_2$ mitigate overestimation bias. The critics are updated with soft Bellman targets
\begin{equation}
   \begin{aligned}
        y = r + \gamma\,\mathbb{E}_{a'\sim\pi_\theta(\cdot|s')}\!\bigl[\min(Q_1(s',a'),Q_2(s',a')) \\ - \alpha\log\pi_\theta(a'|s')\bigr].
   \end{aligned}
\end{equation}
SAC's maximum-entropy objective Eq (\ref{eq:entropy_dsac}) naturally complements the diffusion policy's multi-modality, encouraging broad action coverage while maintaining high Q-value outcomes. DM-enabled SAC has been extended with attention mechanisms to handle multi-service heterogeneous environments \cite{liu2025qos,liu2024towards}, with meta-learning for fast adaptation in multi-UAV systems \cite{zhang2025diffusion}, with digital twins to pre-verify decisions before execution \cite{xiong2025diffusion}, and with transfer learning for domain-adaptive inference acceleration \cite{tian2025accelerating}.

\subsubsection{DM-enabled Twin Delayed Deep Deterministic Policy Gradient}

Twin Delayed Deep Deterministic Policy Gradient (TD3) \cite{fujimoto2018addressing} is an off-policy algorithm designed for deterministic continuous control with a stable twin-critic structure and delayed policy updates. The diffusion variant (DM-enabled TD3 or GDMTD3) replaces the TD3 actor with a DDPM policy while retaining target policy smoothing. The actor gradient is

\begin{equation}
\begin{aligned}
\nabla_\theta J \approx \mathbb{E}_{s\sim\mathcal{D}}\Big[ 
& \nabla_{a_0}\min\big(Q_1(s,a_0), Q_2(s,a_0)\big) \Big|_{a_0=\pi_\theta(s,\xi)}\\ & 
\cdot \nabla_\theta \pi_\theta(s,\xi)
\Big],
\end{aligned}
\end{equation}
where $\xi$ is the noise trajectory used during the reverse denoising chain. Here, a noise $\tilde{\epsilon}\sim\mathrm{clip}(\mathcal{N}(0,\sigma^2),-c,c)$ is injected into denoised target actions to smooth the critic landscape (i.e., target policy smoothing). The DDPM loss in \eqref{loss_elbo} serves as an additional regularizer that prevents the actor from collapsing to a single mode. DM-enabled TD3 is commonly adopted in physical-layer security tasks involving beamforming and trajectory optimization \cite{zhang2024multi,liang2024uav}, as well as in integrated sensing and communication systems \cite{xie2025multi}, where high-precision deterministic control is required in adversarial or constrained environments.

\subsubsection{DM-based Deep Deterministic Policy Gradient}

Deep Deterministic Policy Gradient (DDPG) \cite{lillicrap2015continuous} is a foundational off-policy algorithm for continuous action spaces. The diffusion variant (DM-enabled DDPG or GDMDDPG) integrates a DDPM into the DDPG actor to enable multi-modal action generation in its otherwise deterministic framework. The critic is updated with the standard Bellman loss
\begin{equation}
    L_{\mathrm{critic}}(\phi) = \mathbb{E}\!\left[\bigl(Q_\phi(s,a) - (r + \gamma\, Q_{\phi'}(s',\pi_{\theta'}(s')))\bigr)^2\right],
\end{equation}
where $\phi'$ and $\theta'$ denote target network parameters. The actor maximizes the expected Q-value through policy gradient
\begin{equation}
    \nabla_\theta J = \mathbb{E}_{s}\!\left[\nabla_{a}Q_\phi(s,a)\big|_{a=\pi_\theta(s)}\cdot\nabla_\theta\pi_\theta(s)\right].
\end{equation}
The diffusion denoiser $\epsilon_\theta$ generates $a_0$ via the reverse process \eqref{eq:diffusion_reverse}, and the policy gradient is backpropagated through the final denoised action. DM-enabled DDPG is widely used in MEC offloading problems \cite{cao2024joint} and multi-dimensional resource allocation \cite{xie2025multi}, where the moderate dimensionality of the action space makes it more computationally efficient than full entropy-based formulations.

\subsubsection{DM-based Proximal Policy Optimization}

Proximal Policy Optimization (PPO) \cite{schulman2017proximal} is an on-policy algorithm that constrains each gradient update via a clipped probability ratio, balancing exploration with stability. When a DDPM policy is adopted as the actor, PPO can leverage the expressive multimodal action generation capability of diffusion models while retaining stable policy optimization through clipped updates..Integrating a DDPM actor into PPO requires computing the importance-sampling ratio $r_t(\theta) = \pi_\theta(a|s)/\pi_{\theta_{\mathrm{old}}}(a|s)$. Since the exact likelihood of a diffusion policy is intractable, it is approximated as the product of Gaussian reverse-kernel densities:
\begin{equation}
    \log\pi_\theta(a_0|s) \approx \sum_{k=1}^{T}\log p_\theta(a_{k-1}|a_k, s).
\end{equation}
The clipped surrogate objective is then
\begin{equation}
    L^{\mathrm{CLIP}}(\theta) = \mathbb{E}\!\left[\min\!\left(r_t(\theta)\hat{A}_t,\;\mathrm{clip}\!\left(r_t(\theta),1-\varepsilon,1+\varepsilon\right)\hat{A}_t\right)\right],
\end{equation}
where $\hat{A}_t$ is the generalized advantage estimate. This combination of DMs and PPO enables better modeling of complex action distributions, stronger exploration, and better performance in high-dimensional continuous action spaces. DM-enabled PPO has been applied to uplink resource allocation in LEO satellite networks by combining it with a parameterized action MDP to handle hybrid discrete-continuous decision spaces \cite{wang2025uplink}, and to privacy-preserving offloading where the diffusion denoising process additionally transforms raw states into privacy-preserving representations \cite{you2024generative}. The on-policy nature of PPO makes it suitable for scenarios where the system dynamics change rapidly and maintaining an up-to-date policy is critical.

\subsubsection{Multi-Agent Deep Reinforcement Learning with DM-based Policies}

Several network models involve multiple interacting agents, such as UAVs, vehicles, or edge servers, that must coordinate their decisions under partial observability. The standard framework follows Centralized Training with Decentralized Execution (CTDE): each agent $i$ maintains a local diffusion actor $\pi_{\theta_i}$ that generates actions $a_i$ from local observation $o_i$, while a centralized critic evaluates a joint value function that conditions on the global state.

\noindent\textbf{DM-enabled MADDPG.}
In the multi-agent extension of DDPG, the centralized critic for agent $i$ takes all agents' observations and actions as input: $Q_{\phi_i}(o_1,\ldots,o_N,a_1,\ldots,a_N)$. When diffusion models are integrated into MADDPG, each agent's deterministic actor is replaced by a conditional denoising policy that iteratively generates actions from Gaussian noise conditioned on the local observation. The centralized critic still evaluates the joint action configuration, while the diffusion actor enables each agent to represent richer and potentially multimodal behaviors, improving coordination and exploration in complex multi-agent environments. The actor update is:
\begin{equation}
    \nabla_{\theta_i}J \approx \mathbb{E}\!\left[\nabla_{a_i}Q_{\phi_i}(\mathbf{o},\mathbf{a})\Big|_{a_i=\pi_{\theta_i}(o_i)}\!\cdot\nabla_{\theta_i}\pi_{\theta_i}(o_i)\right],
\end{equation}
where $\mathbf{o}=(o_1,\ldots,o_N)$ and $\mathbf{a}=(a_1,\ldots,a_N)$. Each $\pi_{\theta_i}$ generates its action via the reverse diffusion chain conditioned on $o_i$. This formulation has been used for cooperative task assignment among UAV swarms \cite{tang2025dnn,tang2025task} and for semantic offloading in vehicular edge computing \cite{yang2025diffusion}, where local observations must drive globally coordinated decisions.

\noindent\textbf{DM-enabled QMIX.}
Value decomposition methods such as QMIX \cite{rashid2020monotonic} factorize the joint action-value function into agent-wise utilities via a monotone mixing network $Q_{\mathrm{tot}} = f_{\mathrm{mix}}(Q_1,\ldots,Q_N)$, which ensures consistent greedy action selection. Integrating diffusion actors into QMIX combines the expressive policy generation of diffusion models with the scalable value decomposition of QMIX, enabling efficient decentralized execution in large-scale networks. This combination has been applied to hierarchical spectrum management \cite{ning2025diffusion}, where high-level inter-cluster decisions are generated by a DM-enabled SAC agent while within-cluster allocation uses QMIX-structured local agents.

\noindent\textbf{Diffusion with Mean-Field Approximation.}
When the number of agents $N$ is very large, maintaining a centralized critic becomes computationally intractable. Mean-field MARL addresses this by approximating the influence of all neighbors on agent $i$ through the mean action $\bar{a}_i = \frac{1}{|\mathcal{N}_i|}\sum_{j\in\mathcal{N}_i} a_j$, reducing the joint Q-function to $Q_i(o_i, a_i, \bar{a}_i)$. Combining this with a diffusion-based trajectory planner enables scalable multi-agent coordination in dense networks. The diffusion model generates locally coherent action sequences conditioned on the mean-field observation, while the mean-field approximation avoids full information exchange. This has been demonstrated in multi-agent spectrum planning under both ideal and limited radio-frequency conditions \cite{meng2025multi}, achieving faster convergence and lower packet loss than traditional MARL baselines.

\noindent\textbf{Diffusion with Advantage Actor-Critic.}
In the multi-agent advantage actor-critic (A2C) framework, each agent's actor is updated using the advantage $A(s,a) = Q(s,a)-V(s)$ as a baseline to reduce variance. Replacing the actor with a DM and organizing agents under a distributed orchestration mechanism enables large-scale cloud scheduling, where the A2C centralized critic provides gradient direction to diffusion actors at each edge node \cite{wang2024dmais}. The Lyapunov-constrained variant further stabilizes long-term resource usage by transforming the infinite-horizon problem into per-slot deterministic subproblems, with diffusion actors producing multi-modal offloading decisions at each slot \cite{rao2025computation,liu2024dnn}.

\subsubsection{Diffusion-Based Planning}

Beyond single-step action generation, diffusion models serve as trajectory-level planners, generating entire action sequences as a single coherent diffusion sample. Drawing on the Diffuser framework \cite{Janner2022PlanningWD}, the planner treats a full trajectory $\tau = (s_0,a_0,s_1,a_1,\ldots,s_H,a_H)$ as a high-dimensional data point and learns a diffusion model over trajectory space. The reward-conditioned distribution is obtained via classifier guidance:
\begin{equation}
    p_\theta(\tau|R) \propto p_\theta(\tau)\exp\!\left(\gamma\,\mathcal{R}(\tau)\right),
\end{equation}
where $\mathcal{R}(\tau)=\sum_{t=0}^H r_t$ is the cumulative reward and $\gamma>0$ is the guidance strength. The reward gradient $\nabla_{\tau_k}\mathcal{R}(\tau_k)$ is added to each denoising step, steering the trajectory toward higher returns while $p_\theta(\tau)$ enforces physical and behavioral feasibility. The first action $a_0$ of the generated trajectory is executed, and the planner is called again from the new state.

To improve real-time applicability in edge systems, DiffuserLite \cite{dong2024diffuserlite} adopts a coarse-to-fine refinement strategy: a low-resolution trajectory is generated in few denoising steps and then progressively refined, improving decision frequency by more than two orders of magnitude while maintaining near-optimal performance. Diffusion planners are particularly suited to multi-step vehicle routing \cite{qiao2025combined} and robotic manipulation \cite{li2023crossway,he2023diffusion}, where long-horizon temporal coherence and constraint satisfaction across multiple steps are critical. In multi-task settings, diffusion planners can additionally serve as data synthesizers by generating high-fidelity synthetic transitions to augment offline datasets \cite{he2023diffusion}.

\subsubsection{Diffusion for Data Augmentation}

A complementary use of diffusion models in DRL is to enrich the experience replay buffer with synthesized transitions rather than to parameterize the policy directly. This approach leverages the DM's generative capacity to overcome data sparsity, poor coverage of rare states, and distributional shift between training and deployment environments.

Given a small set of real transitions $\{(s, a, r, s')\}$, a conditional diffusion model $p_\theta(s', r \mid s, a)$ is trained to generate plausible next states and rewards. The generated transitions are mixed with real data in the replay buffer before each batch update:
\begin{equation}
    \mathcal{B}_{\mathrm{aug}} = \mathcal{B}_{\mathrm{real}} \cup \{(s, a, r', s') : (r', s') \sim p_\theta(\cdot\mid s, a)\}.
\end{equation}
This strategy improves sample efficiency and reduces the risk of overfitting to limited operational data. In the Diffusion-enable DQN framework \cite{shi2025diffusion}, the diffusion-based data synthesizer generates synthetic high-quality transitions to enrich the replay buffer, which a trajectory planner then uses to limit multi-step planning errors. The result is at least a 12.5\% reduction in end-to-end delay compared with standard DQN in traffic control scenarios. Similarly, in the GAI-CFL framework \cite{he2025dual}, a diffusion model generates optimized heterogeneous data distributions for each local device to address multi-granularity heterogeneity in clustered federated learning, achieving 60\% higher test accuracy than traditional federated learning. Data augmentation with diffusion models is especially effective in scenarios where real interaction is expensive or hazardous, such as satellite edge computing \cite{rao2025computation} and physical-layer security \cite{zhang2024multi}.

\subsubsection{Diffusion for Reward Shaping}

In sparse-reward environments, standard DRL algorithms converge slowly because meaningful gradient signals are infrequent. Diffusion-based reward shaping addresses this by using a diffusion model to synthesize informative intermediate reward signals that guide the agent toward high-return regions of the state-action space before the true terminal reward is observed.

The DRESS framework \cite{you2025dress} introduces a diffusion reasoning module that conditions on the current state-action pair to infer a dense reward signal $\tilde{r}(s,a)$. This shaped reward is used to augment the environment reward during training:
\begin{equation}
    r^{\mathrm{shaped}}(s,a,s') = r(s,a,s') + \eta\,\tilde{r}(s,a),
\end{equation}
where $\eta>0$ controls the shaping strength. The diffusion model effectively imputes the missing credit signal by reasoning about the long-run consequences of actions in a learned latent space. DRESS achieves approximately 1.5$\times$ faster convergence than baseline DRL in sparse-reward UAV-assisted wireless networks. Beyond explicit reward shaping, the GDM-based intent-guided trajectory generation in \cite{wu2025drl} enables fine-tuning the policy's reward objective online, adapting to target QoS requirements without full retraining and thereby reducing dependence on dense environment feedback.

\begin{table*}[!h]
\caption{Summary of DM-enabled DRL Methods Used in This Survey}
\label{tab:dm_drl_summary}
\centering
\renewcommand{\arraystretch}{1.4}
\resizebox{\textwidth}{!}{%
\begin{tabular}{|l|l|l|l|l|}
\hline
\makecell[l]{\textbf{Method}} &
\makecell[l]{\textbf{Base} \\ \textbf{Algorithm}} &
\makecell[l]{\textbf{Advantages}} &
\makecell[l]{\textbf{Limitations}} &
\makecell[l]{\textbf{Scenarios in This Survey}} \\
\hline
\makecell[l]{\textbf{DM-enabled QL} \\ \cite{wang2022diffusion}} &
\makecell[l]{Offline \\ Q-learning} &
\makecell[l]{Strong behavioral regularization; avoids \\ out-of-distribution actions; no environment \\ interaction required} &
\makecell[l]{Q-function must generalize from static data; \\ limited to offline settings; multiple denoising \\ steps increase inference latency} &
\makecell[l]{Offline spectrum allocation \cite{nouri2025diffusion}; \\ power control \cite{darabi2024diffusion}; \\ offline robot manipulation \cite{wang2025integrating}} \\
\hline
\makecell[l]{\textbf{DM-enabled SAC} \\ \cite{wang2024diffusion}} &
\makecell[l]{SAC} &
\makecell[l]{Multi-modal action generation; entropy \\ regularization promotes exploration; \\ stable online training} &
\makecell[l]{Intractable exact entropy; higher per-step \\ cost than Gaussian SAC; requires \\ tuning temperature parameter} &
\makecell[l]{MEC offloading \cite{du2024diffusion,xu2025enhancing}; \\ AIGC service selection \cite{liu2025qos,liu2024towards}; \\ UAV resource allocation \cite{zhang2025diffusion}; \\ trust-aware consensus \cite{chen2025trust}} \\
\hline
\makecell[l]{\textbf{DM-enabled TD3} \\ \cite{fujimoto2018addressing}} &
\makecell[l]{TD3} &
\makecell[l]{Stable training via twin critics and \\ delayed updates; effective for high-precision \\ deterministic control} &
\makecell[l]{Target policy smoothing needs re-design \\ for stochastic diffusion actors; less expressive \\ than SAC in highly multi-modal tasks} &
\makecell[l]{UAV physical-layer security \cite{zhang2024multi,liang2024uav}; \\ ISAC beamforming \cite{xie2025multi}; \\ vehicular metaverse \cite{tong2024diffusion}} \\
\hline
\makecell[l]{\textbf{DM-enabled DDPG} \\ \cite{gu2016continuous}} &
\makecell[l]{DDPG} &
\makecell[l]{Computationally lightweight; straightforward \\ policy gradient; suitable for moderate- \\ dimensional continuous action spaces} &
\makecell[l]{No entropy bonus; prone to overestimation \\ bias without twin critics; less stable \\ in non-stationary environments} &
\makecell[l]{Joint offloading and resource allocation \cite{cao2024joint}; \\ IRS-assisted systems \cite{xie2025multi}; \\ DNN partitioning in VEC \cite{liu2024dnn}} \\
\hline
\makecell[l]{\textbf{DM-enabled PPO} \\ \cite{schulman2017proximal}} &
\makecell[l]{PPO} &
\makecell[l]{On-policy updates ensure policy freshness; \\ clipped ratio prevents destructive updates; \\ suits fast-changing environments} &
\makecell[l]{Exact likelihood is intractable, requires \\ approximation; higher variance than \\ off-policy methods; sample-inefficient} &
\makecell[l]{Uplink RSMA in LEO networks \cite{wang2025uplink}; \\ privacy-preserving VR-AIGC offloading \cite{you2024generative}; \\ intent-guided wireless control \cite{wu2025drl}} \\
\hline
\makecell[l]{\textbf{DM-enabled MADDPG} \\ \textbf{/ MADRL}} &
\makecell[l]{MADDPG \\ (CTDE)} &
\makecell[l]{Handles partial observability; centralized critic \\ improves credit assignment; each agent \\ generates multi-modal actions} &
\makecell[l]{Centralized critic scales poorly with agent \\ count; non-stationarity from simultaneous \\ policy updates} &
\makecell[l]{UAV task assignment \cite{tang2025dnn,tang2025task}; \\ semantic VEC offloading \cite{yang2025diffusion}; \\ decentralized edge dispatch \cite{peng2025decentralized}; \\ LLM QoS scheduling \cite{yao2025enhancing}} \\
\hline
\makecell[l]{\textbf{DM-enabled QMIX} \\ \cite{rashid2020monotonic}} &
\makecell[l]{QMIX} &
\makecell[l]{Scalable value decomposition; decentralized \\ execution with global optimization; monotone \\ mixing preserves greedy action selection} &
\makecell[l]{Limited to cooperative tasks; cannot represent \\ arbitrary joint Q-functions; mixing network \\ may underfit complex value landscapes} &
\makecell[l]{Hierarchical spectrum allocation \cite{ning2025diffusion}; \\ multi-agent DNN offloading \cite{liu2024dnn}} \\
\hline
\makecell[l]{\textbf{Diffusion with} \\ \textbf{Mean-Field MARL}} &
\makecell[l]{Mean-Field \\ MARL} &
\makecell[l]{Scalable to very large agent populations; \\ avoids full joint observation; captures \\ local neighborhood dependencies} &
\makecell[l]{Loses inter-agent correlations beyond \\ one hop; requires homogeneous \\ agent structure} &
\makecell[l]{Dense spectrum planning \cite{meng2025multi}} \\
\hline
\makecell[l]{\textbf{Diffusion A2C} \\ \textbf{with Lyapunov}} &
\makecell[l]{A2C + \\ Lyapunov} &
\makecell[l]{Advantage baseline reduces gradient variance; \\ Lyapunov constraint ensures long-term \\ stability without global optimization} &
\makecell[l]{On-policy updates are sample-inefficient; \\ Lyapunov design requires system \\ model knowledge} &
\makecell[l]{Large-scale cloud scheduling \cite{wang2024dmais}; \\ satellite edge offloading \cite{rao2025computation}; \\ VEC with DNN partitioning \cite{liu2024dnn}} \\
\hline
\makecell[l]{\textbf{DM-enabled Based} \\ \textbf{Planning} \\ \cite{Janner2022PlanningWD}} &
\makecell[l]{Offline DRL \\ / BC} &
\makecell[l]{Generates temporally coherent trajectories; \\ jointly captures state-action dependencies; \\ enforces constraints across the horizon} &
\makecell[l]{High inference latency per trajectory; \\ memory-intensive for long horizons; \\ unsuited for real-time control without \\ approximation} &
\makecell[l]{Vehicle routing \cite{qiao2025combined}; \\ robotic manipulation \cite{li2023crossway,he2023diffusion}; \\ autonomous driving planning \cite{dong2024diffuserlite}} \\
\hline
\makecell[l]{\textbf{Diffusion for} \\ \textbf{Data Augmentation}} &
\makecell[l]{Any DRL} &
\makecell[l]{Increases sample diversity; reduces overfitting \\ to sparse offline data; requires no \\ policy architecture change} &
\makecell[l]{Hallucinated transitions may mislead \\ Q-functions; requires accurate \\ environment dynamics model} &
\makecell[l]{Traffic routing \cite{shi2025diffusion}; \\ federated edge learning \cite{he2025dual}; \\ multi-task offline DRL \cite{he2023diffusion}} \\
\hline
\makecell[l]{\textbf{Diffusion for} \\ \textbf{Reward Shaping}} &
\makecell[l]{Any DRL} &
\makecell[l]{Resolves sparse reward problems; \\ accelerates convergence; compatible \\ with any policy architecture} &
\makecell[l]{Shaped reward introduces bias if diffusion \\ model is miscalibrated; additional \\ model training overhead} &
\makecell[l]{Sparse-reward UAV networks \cite{you2025dress}; \\ intent-driven wireless scheduling \cite{wu2025drl}} \\
\hline
\makecell[l]{\textbf{Hierarchical DRL} \\ \textbf{with Diffusion}} &
\makecell[l]{Hi: DM \\ Lo: SAC / \\ PPO / IPPO} &
\makecell[l]{Decomposes problems across timescales; \\ high-level diffusion generates abstract goals \\ with multi-modal diversity} &
\makecell[l]{Inter-level coordination requires careful \\ reward decomposition; coupled updates \\ cause non-stationarity} &
\makecell[l]{UAV-UGV cooperative sensing \cite{zhao2024energy}; \\ hierarchical spectrum management \cite{ning2025diffusion}} \\
\hline
\makecell[l]{\textbf{Transfer DRL} \\ \textbf{with Diffusion}} &
\makecell[l]{SAC + \\ Transfer DRL} &
\makecell[l]{Rapid policy adaptation to new environments; \\ few target-domain samples needed; \\ leverages source-domain knowledge} &
\makecell[l]{Performance degrades under large domain \\ gaps; diffusion model training \\ is domain-specific} &
\makecell[l]{AIGC collaborative inference \\ under mobility \cite{tian2025accelerating}} \\
\hline
\end{tabular}%
}
\end{table*}

\subsubsection{Hierarchical and Transfer DRL with Diffusion}

\noindent\textbf{Hierarchical DRL.}
Many real-world systems require decisions at multiple timescales: a high-level planner sets abstract goals (e.g., which cluster to serve, which server to migrate to) while a low-level controller executes fine-grained actions (e.g., beamforming weights, resource block allocation). Hierarchical DRL decomposes this into two interacting agents. In the diffusion variant, the high-level agent is equipped with a diffusion policy that generates abstract subgoals or macro-actions $g\sim\pi_{\mathrm{hi}}(\cdot|s)$ via reverse denoising, while the low-level agent $\pi_{\mathrm{lo}}(a|s,g)$ conditions on both the state and the subgoal. The high-level diffusion actor is trained to maximize the cumulative reward accrued over a horizon of $K$ low-level steps, while the low-level agent maximizes a goal-reaching intrinsic reward. This structure has been applied to cooperative UAV-UGV sensing \cite{zhao2024energy}, where UGVs (Unmanned Ground Vehicles) are navigated by a discrete diffusion high-level policy to rendezvous points optimized by Multi-Agent SAC, achieving nearly double the UAV–UGV cooperation factor compared to flat MARL baselines. It has also been used in hierarchical spectrum allocation \cite{ning2025diffusion}, where a DM-enabled SAC high-level agent allocates inter-cluster spectrum and a QMIX low-level structure handles within-cluster resource blocks.

\noindent\textbf{Transfer DRL with Diffusion.}
Deploying a policy trained in a source environment to a different target environment is challenged by distributional shift in dynamics, channel statistics, or user density. A diffusion model can bridge this gap by learning a domain-adaptive transition distribution. In the adaptive forward-predict-reflect (AFPR) mechanism \cite{tian2025accelerating}, a diffusion model is trained to generate policy actions robust to domain shifts: the forward process encodes source-domain policy behaviors into a latent trajectory, and the reverse process reconstructs target-domain actions conditioned on updated environmental observations. This enables the policy to adapt to new deployment conditions with few target-domain samples, significantly reducing response latency and improving system utility in dynamic edge-assisted AIGC inference scenarios.

\subsection{Summary of DM-enabled DRL Methods}

Table~\ref{tab:dm_drl_summary} summarizes the DM-enabled DRL methods discussed above, highlighting the base DRL algorithm, key advantages, main limitations, and the representative application scenarios from this survey.

\subsection{Discussion}

The methods reviewed in this section represent a rich design space for integrating diffusion models into deep reinforcement learning. Several overarching observations emerge from their collective properties and usage patterns in the surveyed literature.

\noindent\textbf{Prevalence of actor substitution.}
The most frequently adopted integration strategy is to substitute the conventional Gaussian actor with a DDPM policy within an existing actor-critic framework such as SAC, TD3, or DDPG. This approach requires only modest modifications to established algorithms, the critic and replay buffer remain unchanged, while conferring the full multi-modal expressiveness of diffusion sampling on the policy. The widespread adoption of DM-enabled SAC in particular reflects its favorable balance between entropy-driven exploration, twin-critic stability, and compatibility with the continuous action spaces common in resource management problems.

\noindent\textbf{Offline versus online settings.}
DM-enabled QL is tailored to offline settings where environment interaction is prohibited or costly, such as power control with sensitive infrastructure. Its behavioral regularization is essential for avoiding out-of-distribution actions that a Q-function trained on a static dataset cannot reliably evaluate. By contrast, DM-enabled SAC, DM-enabled TD3, and DM-enabled PPO operate in the online regime and are preferred when the system is accessible for exploration, such as MEC offloading or UAV scheduling. The distinction between offline and online deployment must therefore be a primary design consideration when selecting a DM-enabled enabled DRL method.

\noindent\textbf{Multi-agent scalability.}
The extension of diffusion policies to multi-agent settings spans a spectrum of scalability–optimality trade-offs. Full CTDE with DM-enabled MADDPG achieves the best credit assignment but becomes computationally intractable for large agent populations. QMIX-based decomposition offers a scalable middle ground for cooperative tasks, while mean-field approximations enable deployment in dense networks at the cost of losing long-range inter-agent correlations. Selecting among these architectures requires careful consideration of the system size, degree of agent coupling, and available communication infrastructure.

\noindent\textbf{Computational cost and latency.}
A recurring limitation across all DM-enabled DRL methods is the multi-step denoising inference required to generate each action or trajectory, which increases latency relative to a single forward pass through a Gaussian policy. For near-real-time resource allocation tasks, this cost can be mitigated through consistency distillation, reduced denoising steps (typically $T=5$–$20$ in practice rather than $T=1000$ used in image generation), or coarse-to-fine trajectory refinement as in DiffuserLite. The computational overhead is most acute in planning-based methods operating over long horizons, and least problematic in single-step actor substitution methods where GPU-level parallelism makes denoising fast enough for most network control timescales.

\noindent\textbf{Complementary roles of diffusion.}
Beyond policy parameterization, diffusion models play complementary roles in DRL pipelines: as data synthesizers that enrich replay buffers \cite{shi2025diffusion,he2025dual}, as reward-shaping modules that resolve sparse feedback \cite{you2025dress}, and as domain-adaptation bridges in transfer DRL \cite{tian2025accelerating}. These auxiliary roles can be combined with any of the core actor-critic methods above, offering practitioners a modular toolkit for addressing the specific bottlenecks-data scarcity, reward sparsity, or distributional shift-present in their target application. The choice of which role diffusion should play-policy, planner, augmenter, or shaper-ultimately depends on whether the primary challenge is representational expressiveness, sample efficiency, convergence speed, or generalization.

\section{Computation Offloading}
\label{sec:computation_offloading}
In mobile edge computing (MEC), UAV-assisted networks, vehicular edge systems, and AIGC-driven services, computation offloading faces distinct scenario-specific challenges. These include resource heterogeneity, user mobility, time-varying channels, dynamic task arrivals, stringent latency requirements, and diverse service demands. Such factors make offloading decisions high-dimensional, non-convex, and strongly coupled. A single system state may correspond to multiple feasible actions, such as local execution, edge offloading, cooperative processing, service migration, or model selection, leading to multi-modal and discontinuous decision boundaries. Conventional RL methods often suffer from limited policy expressiveness, inefficient exploration, slow convergence, and unstable training in these complex environments. In contrast, diffusion models generate actions via iterative denoising and can more effectively capture complex action distributions. As a result, DM-enabled DRL improves exploration, enhances policy expressiveness, and stabilizes learning, making it suitable for robust offloading and resource coordination in heterogeneous edge intelligence systems. In this section, we review existing DM-enabled DRL schemes for computation offloading, organized by specific application scenarios, as summarized in Table \ref{diffusion_co_summary}.

\begin{table*}[t]
\centering
\caption{Summary of DM-enabled Computation Offloading}
\label{diffusion_co_summary}
\renewcommand{\arraystretch}{1.5}

\begin{tabular}{|l|l|l|l|l|l|}
\hline
\textbf{Category} & \textbf{Ref.} & \textbf{Scenario} & \textbf{Objective} & \textbf{Method} & \textbf{Diffusion Role} \\
\hline

\multirow{6}{*}{\makecell[l]{MEC Offloading\\ \& Resource Allocation}}
& \cite{du2024integrated} & WiFi MEC & Cost Minimization & DM-enabled TD3 & Action Generation \\ \cline{2-6}
& \cite{rao2025computation} & Satellite MEC & Delay Minimization & DM-enabled SAC & Policy Stabilization \\ \cline{2-6}
& \cite{cao2024joint} & IoT MEC & Latency Minimization & DM-enabled DDPG & Policy Modeling \\ \cline{2-6}
& \cite{wang2025energy} & Low-Altitude MEC & Energy Efficiency Maximization & DM-enabled DRL & Action Generation \\ \cline{2-6}
& \cite{peng2025decentralized} & Edge-Cloud MEC & QoS Maximization & DM-enabled MARL & Coordinated Policy \\ \cline{2-6}
& \cite{xu2025enhancing} & Collaborative Edge & QoE Maximization & DM-enabled SAC & Exploration Enhancement \\
\hline

\multirow{6}{*}{\makecell[l]{UAV-Assisted\\ Offloading \& Scheduling}}
& \cite{tong2024diffusion} & UAV Metaverse & Migration Cost Minimization & DM-enabled DRL & Decision Generation \\ \cline{2-6}
& \cite{zhang2025priority} & Multi-UAV AIGC & Utility Maximization & DM-enabled DRL & Action Generation \\ \cline{2-6}
& \cite{zhang2025diffusion} & UAV Edge Computing & Reward Maximization & DM-enabled SAC & Policy Generation \\ \cline{2-6}
& \cite{tang2025dnn} & UAV Network & Delay Minimization & Diffusion-MADDPG & Actor Enhancement \\ \cline{2-6}
& \cite{you2025dress} & Intelligent Network & Reward Maximization & DM-enabled DRL & Reward Enhancement \\ \cline{2-6}
& \cite{tang2025task} & UAV Rescue & Latency \& Energy Minimization & DM-enabled MADDPG & Structured Decision \\
\hline

\multirow{5}{*}{\makecell[l]{Vehicular Edge\\ Computing \& Offloading}}
& \cite{yang2025diffusion} & Vehicular MEC & Utility Maximization & DM-enabled MARL & Policy Learning \\ \cline{2-6}
& \cite{liu2024dnn} & Vehicular Network & Cost Minimization & DM-enabled QMIX & Decision Generation \\ \cline{2-6}
& \cite{zhong2025generative} & Vehicular AI & Utility Maximization & DM-enabled DRL & Strategy Generation \\ \cline{2-6}
& \cite{he2025dual} & Federated Edge & Energy Minimization & DM-enabled DRL & Data Augmentation \\ \cline{2-6}
& \cite{huang2024adaptive} & Video Streaming & QoE Maximization & DM-enabled DRL & Exploration Enhancement \\
\hline

\multirow{5}{*}{\makecell[l]{AIGC \& LLM\\ Service Offloading}}
& \cite{du2024diffusion} & Edge AIGC & Utility Maximization & DM-enabled SAC & Policy Generation \\ \cline{2-6}
& \cite{liu2025qos} & Multi-AIGC & Utility Maximization & DM-enabled SAC & Policy Modeling \\ \cline{2-6}
& \cite{liu2024towards} & Multi-Task AIGC & Cost Minimization & DM-enabled SAC & Policy Generation \\ \cline{2-6}
& \cite{tian2025accelerating} & Edge Inference & Latency Minimization & DM-enabled SAC & Decision Support \\ \cline{2-6}
& \cite{yao2025enhancing} & Cloud-Edge LLM & QoS Maximization & DM-enabled MARL & Scheduling Policy \\
\hline

\multirow{2}{*}{\makecell[l]{Edge-Cloud Scheduling\\ \& Network Design}}
& \cite{wang2024dmais} & Edge-Cloud System & Latency Minimization & DM-enabled A2C & Policy Generation \\ \cline{2-6}
& \cite{huang2023ai} & Intelligent Network & Utility Maximization & DM-enabled DRL & Structure Generation \\
\hline

\end{tabular}
\end{table*}

\subsection{MEC Offloading \& Resource Allocation}
MEC offloading and resource allocation involve tightly coupled decisions over computation, communication, and task execution. The optimization problem is usually high-dimensional and non-convex due to heterogeneous edge resources, dynamic task arrivals, and limited wireless bandwidth. Conventional DRL may suffer from inefficient exploration and unstable convergence in such coupled action spaces. DM-enabled DRL is therefore suitable, as it can generate more expressive action distributions and improve policy robustness for joint offloading and resource allocation.

\begin{figure}[t]
    \centering
    \includegraphics[width=\linewidth]{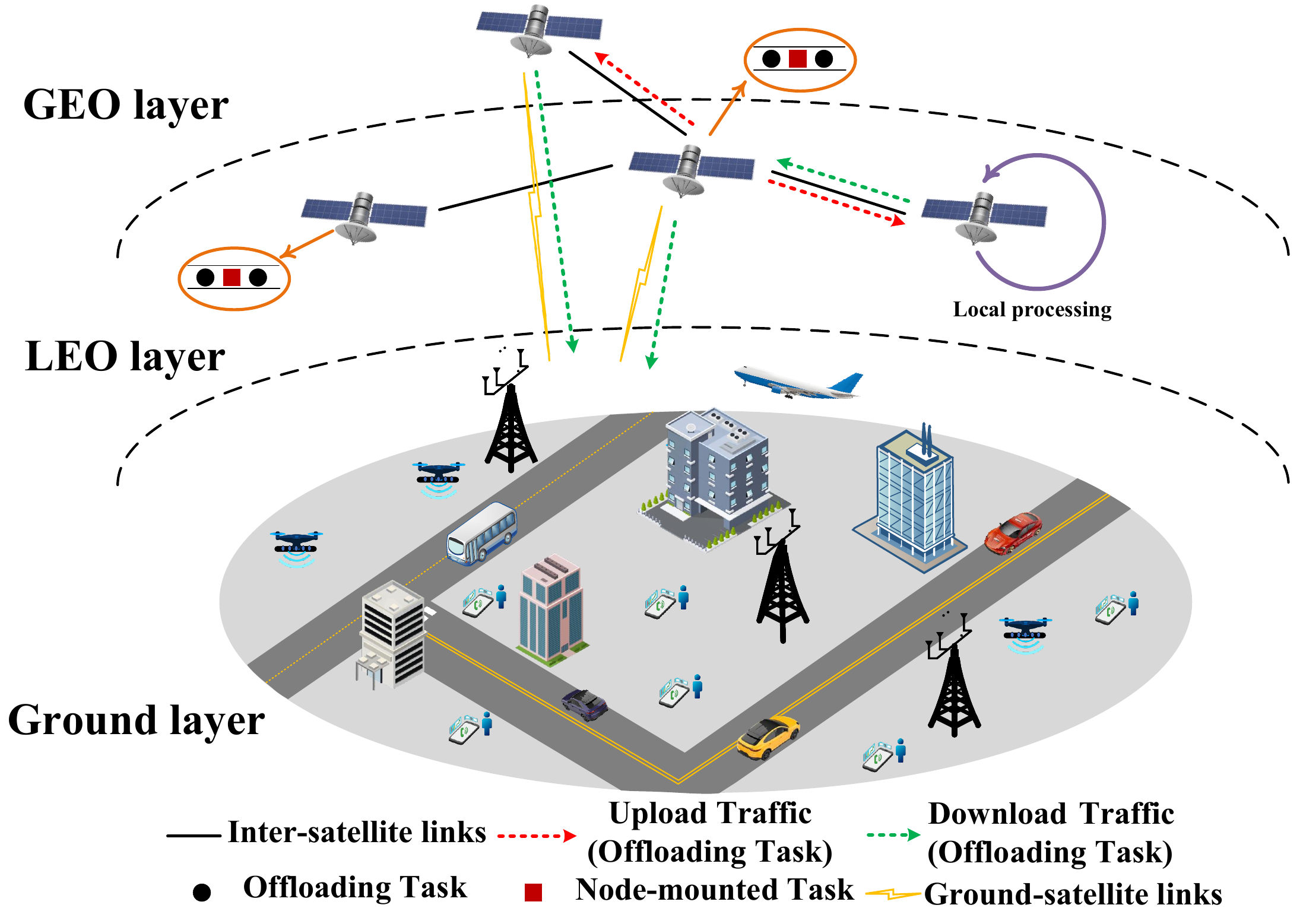}
    \caption{Illustration of multi-satellite cooperative computation offloading systems \cite{rao2025computation}.}
    \label{fig:co}
    \vspace{-2em}
\end{figure}

\begin{figure}[htbp]
    \centering
    \includegraphics[width=\linewidth]{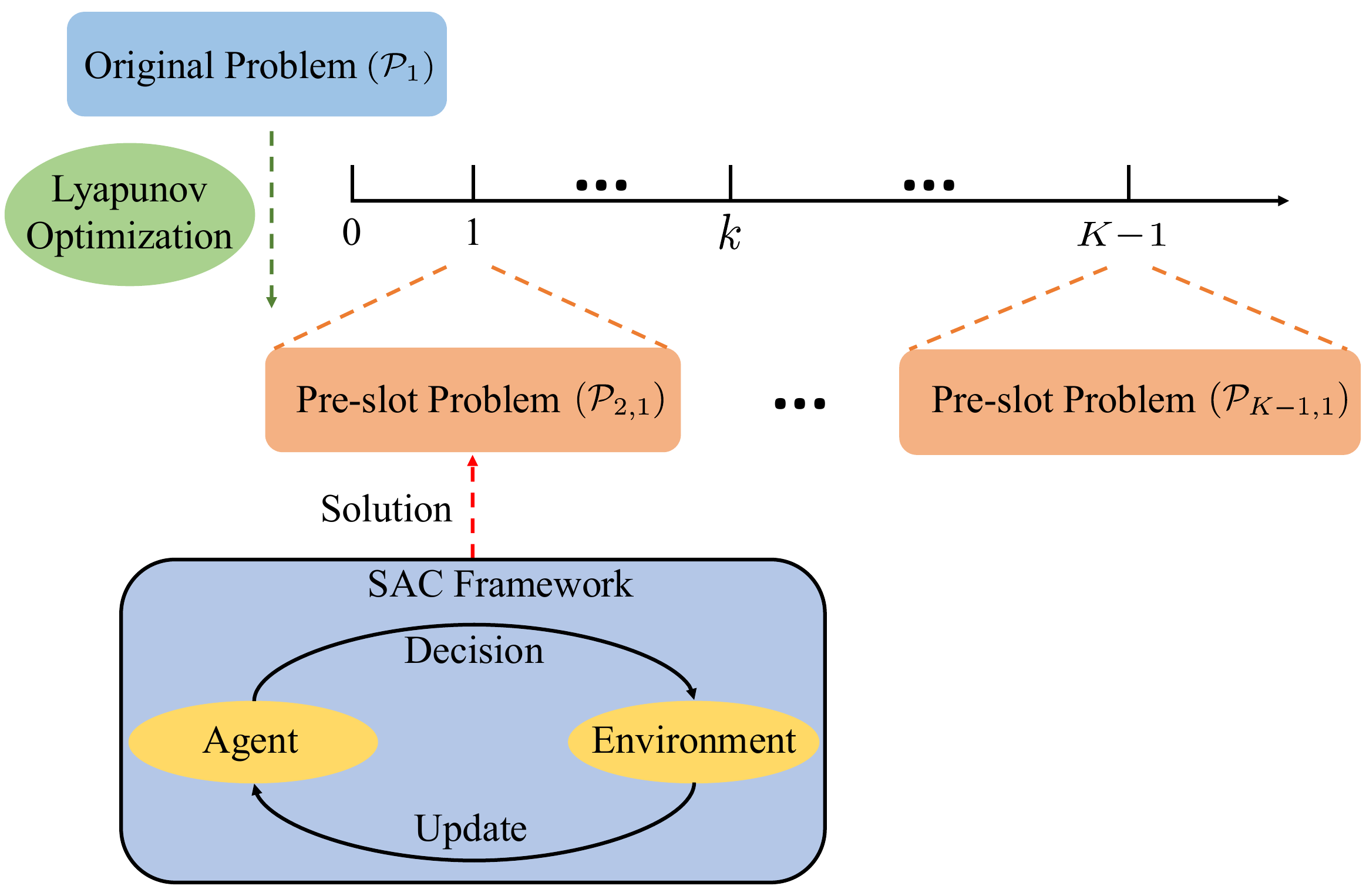}
    \caption{The diagram of the proposed GenAI-DRL scheme in \cite{rao2025computation}.}
    \label{fig:gen}
\end{figure}

The authors in \cite{du2024integrated} consider a joint optimization problem for MEC-enabled IEEE 802.11ax Wi-Fi systems. The formulation includes offloading decisions and resource allocation. A DM-enabled TD3 framework is proposed to enhance policy expressiveness and address sparse-sample limitations in integrated communication-computing environments. The resource allocation subproblem is solved using the Hungarian algorithm, which exploits the matching structure between users and MEC servers. Simulation results show that the proposed method reduces training cost and improves both communication success rate and total cost by up to $100$\% and $33.3$\%, respectively, compared with the benchmarks. However, the method primarily focuses on short-term decision quality, whereas the long-term stability of dynamic edge systems is underemphasized. This motivates the use of stability-aware learning frameworks. Along this line, a satellite edge computing (SEC) system with cooperative multi-SEC processing is examined in \cite{rao2025computation}, as shown in Fig. \ref{fig:co}, where satellite nodes provide heterogeneous computation and communication resources for offloading global tasks. Moreover, a dynamic offloading problem that accounts for diverse task types, time-varying resources, and long-term system sustainability is formulated. As shown in Fig. \ref{fig:gen}, a DM-enabled SAC framework is used to generate high-quality action samples and improve policy expressiveness in high-dimensional SEC environments. The proposed approach also incorporates Lyapunov optimization to transform the long-term offloading objective into an online per-slot problem while maintaining system stability under fluctuating task arrivals. Simulation results demonstrate $37.5$\% improvements in delay reduction and $23.4$\% improvements in system stability compared with existing baselines. Nevertheless, this design depends on accurate queue modeling and may suffer from model mismatch in practical environments. 

 A related direction is to simplify the learning formulation while retaining the ability to make adaptive decisions. Therefore, in \cite{cao2024joint}, an MEC system integrated with IoT is examined, in which users offload computation tasks to edge servers to meet low-latency and high-performance requirements. The authors formulate a joint problem of computing offloading and resource allocation to minimize the total delay across multiple users and servers. A diffusion model is employed as the policy network to generate high-quality action samples in dynamic environments. Based on this design, a DM-enabled DDPG algorithm is developed to learn offloading and allocation decisions jointly. Simulation results show that the proposed scheme achieves a $39.1$\% delay reduction and an $11.3$\% average reward improvement compared to conventional solutions. However, convergence guarantees and performance bounds remain insufficiently discussed.

Compared with the previous delay-oriented studies, energy-aware optimization has received increasing attention. In this direction, the authors in \cite{wang2025energy} propose a rate-splitting multiple access (RSMA)-enabled low-altitude MEC system supported by an UAV, where the UAV assists ground terminals in computation offloading tasks over a shared uplink. The authors formulate a joint optimization problem that involves the UAV's three-dimensional trajectory, RSMA decoding order, task offloading decisions, and resource allocation, aiming to maximize energy efficiency in the presence of uplink interference. A generative DM-enabled SAC network is proposed to generate high-quality actions and enhance exploration in hybrid action spaces. A priority-based RSMA decoding strategy is also introduced to enable efficient successive interference cancellation with low complexity. Simulation results show that the proposed approach outperforms baseline schemes and that integrating generative diffusion modeling with RSMA yields significant energy-efficiency gains. However, the diffusion-assisted design may introduce additional computational overhead at the learning stage.

 To improve scalability in larger edge-cloud environments, a decentralized request dispatch framework for edge-cloud systems is investigated in \cite{peng2025decentralized}, where requests arriving at edge access points are scheduled to either edge nodes or cloud servers based on latency requirements. The authors design a DM-enabled MARL scheme to generate discrete dispatch decisions conditioned on local observations. A QoS-aware resource coordination mechanism is incorporated to proactively filter infeasible actions based on remaining resources and request deadlines, improving the validity of sampled decisions. Experimental results using real traces demonstrate that the proposed approach enhances system throughput by up to 20.1\% compared with representative baselines under realistic dynamic workloads. The results highlight that diffusion policies enhance action expressiveness in MADRL-based scheduling and provide stable performance under stochastic edge-cloud environments. Nevertheless, coordination overhead among agents is not fully characterized.

Building on the need for more stable exploration, the authors of \cite{xu2025enhancing} propose a feedback DM-enabled generative scheduling method to enhance user QoE in collaborative edge systems. The authors incorporate an innovative Feedback Diffusion model into SAC, constructing an effective and efficient framework for task scheduling. Simulation results show that the proposed method reduces service delays by 45.42\% to 87.57\% and speeds up training episode durations by 2.5 times, achieving higher QoE than benchmark methods. However, its reliance on historical action distributions may become less effective when the environment changes rapidly.

\subsection{UAV-Assisted Offloading \& Scheduling}
UAV-assisted offloading introduces strong mobility, dynamic topology, and time-varying air-to-ground channels. The decision process often includes trajectory planning, task assignment, and resource allocation, which makes the action space highly complex. Traditional DRL may struggle with partial observability and unstable learning in rapidly changing environments. DM-enabled DRL can enhance exploration and generate diverse candidate actions, making it effective for robust UAV coordination and adaptive offloading.

The authors in \cite{tong2024diffusion} investigate a vehicle-twin (VT) migration strategy supported by UAVs in an edge-enabled vehicular metaverse, where aerial edge servers assist roadside units (RSUs) to maintain low-latency VT services. The dynamic VT pre-migration and load-balancing tasks are formulated as a sequential decision process to sustain immersive metaverse experiences while mitigating RSU congestion. A DM-enabled SAC approach is employed to generate expressive joint decisions in high-mobility and rapidly changing system states. A heuristic search-based UAV path-planning mechanism is further employed to expand RSU coverage and intelligently alleviate localized overload. Simulation studies verify that the proposed solution achieves faster reward convergence and significantly enhances service reliability and migration responsiveness compared with existing benchmarks. However, the framework primarily focuses on single-agent decision-making and does not fully capture interactions among multiple UAVs. This naturally motivates extending to multi-agent scenarios. In this direction, the priority-aware multi-UAV-aided metaverse system for supporting diverse and time-sensitive AIGC services is investigated in \cite{zhang2025priority}. Specifically, the system is modeled as an MDP that captures the dynamics of UAV operations and task priorities. A DM-enabled DRL algorithm is conceived to optimize resource allocation under heterogeneous service requirements. The simulation results show that the proposed scheme achieves a $16$\% higher reward per step than the SAC counterpart. Nevertheless, the coupling among UAV resources and communication constraints is not explicitly modeled, which may limit performance in dense networks.

To further address joint decision-making, \cite{zhang2025diffusion} investigates a multi-UAV-assisted, edge-enabled metaverse system to support AaaS, where UAVs act as mobile content servers and cooperative relays for ground users. The joint offloading and AIGC resource allocation task is formulated as an MDP that maximizes user utility while accounting for heterogeneous priorities and computational costs. A DM-enabled SAC is introduced by leveraging the reverse diffusion process to generate expressive action samples and enhance long-term decision robustness. The proposed framework is shown to achieve better convergence across various task densities and to maintain high rewards in large-scale dynamic environments. Extensive simulations further indicate that Meta-DSAC achieves superior performance compared with PPO, SAC, and heuristic baselines in terms of long-term reward, service quality, and task completion rate in multi-UAV edge-enabled metaverse scenarios. While this improves policy expressiveness, the learning process remains sensitive to environmental uncertainty and partial observability. This challenge becomes more pronounced in large-scale systems, motivating the integration of multi-agent learning mechanisms. Accordingly, in \cite{tang2025dnn}, a collaborative DNN task-assignment framework for UAV edge networks is proposed, in which multiple UAVs share their computing capabilities to support mobile users. The decision-making process is modeled as a sequential optimization problem that aims to shorten task execution time while respecting resource and mobility constraints. A DM-enabled MARL architecture is proposed, which enables each UAV to produce higher-fidelity actions and adapt more effectively to dynamic operational states. Coordination among UAV agents is realized through an actor-critic learning paradigm with centralized guidance during training. Experimental studies confirm notable improvements in delay reduction, convergence behavior, and overall task execution efficiency when compared with representative MARL approaches. However, the multi-agent training process introduces greater computational complexity and may face scalability issues.

To improve learning efficiency in such complex settings, the authors in \cite{you2025dress} propose a diffusion-reasoning-based reward-shaping scheme for robust network optimization, leveraging diffusion models' reasoning capabilities to infer reward signals from network states and actions. Experimental results indicate that the proposed framework converges approximately 1.5 times faster than the original DRL approach in sparse-reward wireless environments. Although effective, this approach depends on carefully designed reward functions and may lack generality across different scenarios.

Building on the need for more structured and stable decision-making, with the emergence of AI-enabled UAVs and ground-embedded robots (GERs), a cooperation framework involving UAVs, GERs, and airships is proposed in \cite{tang2025task}. This framework formulates the multi-objective task assignment and exploration problem for UAVs as a dynamic, long-term optimization problem. A DM-enabled MARL is incorporated in a Hungarian algorithm. Simulation results demonstrate the effectiveness of the proposed approach, with a significant latency reduction (25\%-30\%) compared to benchmark methods. While this hybrid design improves performance, it increases algorithmic complexity and may limit real-time applicability in highly dynamic UAV environments.

\subsection{Vehicular Edge Computing \& Offloading}
Vehicular edge computing is characterized by high mobility, intermittent connectivity, and stringent latency constraints. Offloading decisions must adapt to fast-changing channels, vehicle positions, and task requirements. Standard DRL policies may converge to suboptimal decisions when the environment changes rapidly. DM-enabled DRL can better model multi-modal decisions and improve policy generalization, which is beneficial for vehicular task offloading and resource scheduling.

A semantic vehicular edge computing (VEC) system with multi-agent coordination is investigated in \cite{yang2025diffusion}, in which vehicles collaboratively process semantic-aware tasks via dynamic wireless connections. Moreover, the authors address a long-term optimization problem that couples semantic offloading, task scheduling, and distributed resource usage in highly mobile vehicular environments. A DM-enabled MADRL framework is introduced to generate expressive joint actions across multiple vehicles and enhance the stability of cooperation under rapidly changing topologies. Furthermore, semantic features are incorporated into the state and action design to reduce computation overhead and improve communication efficiency. Experimental evaluations under realistic vehicular mobility traces indicate that the proposed method achieves significant latency reduction and semantic-aware utility gains compared to traditional MARL baselines. Specifically, the average latency can be reduced by at most $56.3$\% compared to the greedy counterpart when employing $30$ vehicles. However, the method mainly focuses on semantic-aware decision-making, while fast time-varying channel conditions and mobility-induced uncertainty are less explicitly characterized. This motivates more dynamic formulations for vehicular offloading. Along this line, a VEC system supporting DNN tasks is investigated in \cite{liu2024dnn}, where vehicles collaboratively process computation-intensive workloads via V2V and vehicle-to-infrastructure communications. The authors formulate a joint optimization problem that jointly optimizes DNN partitioning, task offloading, and resource allocation to minimize task completion time while maintaining long-term system stability. Moreover, Lyapunov optimization is employed to decompose the long-term problem into a sequence of per-slot deterministic subproblems. A DM-enabled MARL algorithm is then developed, in which DMs make partitioning and offloading decisions in the multi-vehicle environment. Convex optimization techniques are incorporated into MARL as a subroutine for efficient resource allocation. It is demonstrated that the proposed solution achieves $54.59$\% better performance than the greedy algorithm and $21.23$\% better performance than the genetic algorithm in terms of task completion time. Nevertheless, its performance depends on queue dynamics and system-state modeling, which may be difficult to maintain accurately in highly mobile environments.

Beyond DNN partitioning, digital twin migration provides another important vehicular offloading scenario. In this direction, the authors in \cite{zhong2025generative} construct a multidimensional contract-theoretic model between autonomous vehicles and alternative roadside units. A GDM-based algorithm is employed to determine optimal contract designs that enhance the migration efficiency of embodied agent AI twins. Numerical results demonstrate that the convergence speed of GDM is much faster than that of the SAC algorithm, and GDM can achieve near-optimal rewards within a small number of epochs. However, this method focuses more on strategy generation within a contract-theoretic model than on online adaptive learning. This limitation motivates broader system-level designs that jointly consider data, resource, and task heterogeneity. Accordingly, the authors in \cite{he2025dual} propose a dual-circulation generative artificial intelligence (GAI) framework for clustered federated learning CFL to address multigranularity heterogeneity encompassing data, resource, and task heterogeneity. Generative diffusion models are integrated into the GAI-CFL framework to dynamically generate optimal strategies tailored to the characteristics of the input data. Based on the GDMs, a mixture-of-experts method is proposed to select the best expert models. Experimental results show that GAI-CFL achieves a 60\% higher test accuracy than traditional FL in single-heterogeneous data scenarios. For energy optimization of MoE, the configuration with 4 experts and 32-dimensional hidden layers reduces energy consumption by 30\% compared to full-expert activation while maintaining accuracy. Nevertheless, the framework mainly emphasizes data generation and resource optimization, while the direct coupling between DM-enabled policy learning and real-time offloading remains limited.

To further enhance user-centric service quality, the authors in \cite{huang2024adaptive} propose generative AI-based DRL algorithms that are integrated into an adaptive digital twin (DT)-aided communication, computing, and buffer control (3C) management framework to improve action exploration performance. Experimental simulations demonstrate that the proposed DT-assisted 3C management scheme outperforms conventional benchmark methods in terms of QoE, with improvements of 18.4\% and 20.5\% under low and high user dynamics, respectively. However, increased model complexity and digital-twin maintenance overhead may limit scalability in large-scale vehicular or mobile-edge networks.

\subsection{AIGC \& LLM Service Offloading}
AIGC and LLM services introduce computation-intensive and heterogeneous workloads at the edge. Offloading decisions are no longer limited to task placement, but also involve service selection, model selection, inference routing, and quality-aware scheduling. These decisions often show multi-modal and discontinuous structures. DM-enabled DRL can capture complex action distributions and enhance exploration, making it well-suited for adaptive service orchestration in AIGC- and LLM-driven edge systems.

A DM-enabled AIGC service architecture is examined in \cite{du2024diffusion}, in which AIGC-as-a-service is deployed on wireless edge nodes to support metaverse content generation. The system focuses on selecting suitable AIGC service providers (ASPs) for personalized user tasks under uncertain and dynamic network conditions. Specifically, ASP selection is formulated as a sequential decision problem, and a generative, DM-based AI is conceived to capture multimodal decision structures. Moreover, a DM-enabled SAC algorithm is proposed to strike expressive action generation and improved exploration. Simulation results show that the proposed solution outperforms other baselines by at least $25$\% in test reward and offers a scalable approach to AIGC-driven service optimization in wireless networks. However, the service selection problem is primarily formulated within a relatively fixed AIGC service environment, whereas dynamic multi-service coordination is underemphasized. This motivates the development of more flexible orchestration frameworks for multiple AIGC services. Along these lines, in \cite{liu2025qos}, a multi-AIGC service orchestration framework is proposed for resource-constrained edge networks, where mobile users access diverse AIGC services across multiple edge nodes. The authors formulate a QoS-aware orchestration problem and design a user utility function that reflects service capabilities and real-time performance. A DM-enabled SAC algorithm is derived by employing an attention-augmented DM as the policy network to represent complex probability distributions induced by heterogeneous services and dynamic user demands. The attention mechanism extracts task-relevant contextual information to enhance decision quality in multi-service coordination. Experimental results show that the ADSAC algorithm improves the overall user utility by at least $30.4$\% and reduces the server crash rate by at least $17.2$\%. This paper investigates an AIGC service selection mechanism to address diverse user demands in edge environments, where resource-limited edge nodes must respond to heterogeneous content-generation requests. Nevertheless, the method mainly focuses on service orchestration, and the adaptation to diverse task types remains insufficiently explored. To address this issue, a DM-enabled SAC algorithm is introduced in \cite{liu2024towards}, which employs a DM as the policy network within an off-policy RL framework to characterize intricate dependencies between task attributes and edge system states. An attention module is incorporated to capture long-range contextual features and refine the selection process. It is demonstrated that the proposed ADSAC algorithm outperforms existing methods, reducing the overall user utility loss and the server crash rate by at least $58.3$\% and $58.4$\%, respectively. However, its performance still depends on the predefined task-service matching structure, which may limit flexibility under unseen workloads.

Beyond service and model selection, collaborative inference has also become an important direction for AIGC offloading. In this context, a collaborative inference framework for AIGC services in dynamic edge networks is addressed in \cite{tian2025accelerating}, where multiple mobile devices request AIGC outputs from edge servers. The authors model the inference acceleration problem as a sequential decision process that considers device energy consumption, wireless dynamics, and service delay. In this paper, a DM-enabled SAC scheme is designed to render fast adaptation under dynamic edge mobility. Specifically, a generative diffusion model is integrated into a transfer DRL scheme to produce high-quality policy actions robust to domain shifts between training and deployment environments. The proposed adaptive framework enables policy transfer from a source domain to a target domain via a forward-predict-reflect mechanism, thereby improving sample efficiency and maintaining inference performance under dynamic mobility conditions. However, diffusion is mainly involved in the inference model itself, while its role in the offloading policy is less central. This motivates the use of DM-enabled decision frameworks for larger generative service systems. Accordingly, a vector database-assisted cloud-edge collaborative framework for LLM services is proposed in \cite{yao2025enhancing}, in which edge nodes cache LLM responses to reduce latency and costs. The authors design a vector database-assisted cloud-edge LLM quality of service optimization framework that supports general LLMs without modifying their internal architectures. The quality of service optimization task is formulated as an MDP, and a DM-enabled MARL algorithm is developed to determine whether an incoming request should be served by the edge cache or by a cloud LLM. A DM-enabled policy network is introduced to extract request features and generate expressive actions under diverse query patterns. The proposed method is implemented in a real edge system, and experimental results show that the diffusion-enhanced MARL algorithm significantly improves user satisfaction while reducing latency and resource consumption compared to benchmarks.

\subsection{Edge-Cloud Scheduling \& Network Design}
Edge-cloud scheduling and network design require large-scale coordination across distributed edge and cloud resources. The system must handle uncertain request arrivals, service placement, and cross-layer resource allocation. Conventional DRL may face scalability issues and slow convergence in such large decision spaces. DM-enabled DRL provides stronger generative policy modeling and can improve scheduling efficiency, making it useful for system-level orchestration and adaptive network design. However, the framework mainly focuses on policy optimization, while system-level constraints such as communication overhead and cross-layer interactions are not explicitly modeled. This motivates more holistic designs that incorporate higher-level system objectives.

Along this direction, an edge-cloud system with large-scale service scheduling demands is proposed in \cite{wang2024dmais}, where the integration of IoT increases system complexity and decision-making requirements. The authors formulate a scheduling problem to improve system throughput and reduce end-to-end latency under dynamic, large-scale cloud resource conditions. A DM-enabled MARL algorithm is developed, which employs a DM as the policy network within an advantage A2C framework to accelerate policy learning. Moreover, a distributed orchestration mechanism based on multi-agent A2C is introduced to flexibly manage extensive cloud resources. Experimental results show that the proposed solution achieves higher throughput, lower scheduling latency, and faster convergence compared with the baselines.

Based on a diffusion-model-based learning approach, the authors in \cite{huang2023ai} propose an AI-generated network (AIGN) that can automatically obtain solutions in dynamically time-varying environments. Leveraging the advantages of DMs, AIGN exhibits substantial potential to learn reward-maximizing trajectories, autonomously satisfy multiple constraints, adapt to diverse objectives and scenarios, and even intelligently devise novel designs and mechanisms that remain unexplored in existing network environments. Simulation results demonstrate that AIGN converges significantly faster than online algorithms, which require real-time interaction with the environment. Its efficiency is comparable to that of offline batch-constrained deep Q-learning algorithms. Nevertheless, the framework primarily operates in an offline or semi-static setting and lacks adaptive mechanisms for real-time network dynamics. As a result, integrating DM-enabled generative design with online learning and dynamic scheduling remains an open problem in edge-cloud systems.

\subsection{Lessons Learned}
This section reviews works showing that DM-enabled DRL is most useful when computation offloading involves high-dimensional, coupled, and multi-modal decisions. Across MEC, UAV-assisted, vehicular, AIGC/LLM, and edge-cloud scenarios, diffusion models are mainly used to improve policy expressiveness, exploration capability, and training stability. This advantage is especially clear when offloading decisions are jointly optimized with resource allocation, service selection, trajectory control, or scheduling. Another clear trend is the shift from single-agent decision-making to multi-agent and hybrid optimization frameworks. In dynamic scenarios, such as UAV and vehicular networks, diffusion models are often combined with MARL, Lyapunov optimization, matching algorithms, or reward shaping to improve coordination and stability. For AIGC and LLM services, DM-enabled DRL further supports service orchestration, model selection, and inference scheduling under heterogeneous workloads.

\section{Resource Allocation, Security, and Routing}
\label{sec:resource_security_routing}

DM-enabled DRL provides a new paradigm for wireless decision-making by combining the generative modeling capability of diffusion models with the adaptive control capability of DRL. 
In resource allocation, security, and routing-related tasks, DMs are mainly used to generate high-quality actions, trajectories, policies, or state representations, thereby improving exploration, robustness, and generalization in dynamic wireless environments. 
This section reviews recent studies from four perspectives: power allocation, spectrum allocation, security and privacy, and robotics control, vehicle routing, and UAV planning, as summarized in Table~\ref{Tab_Sec_IV}.

\begin{table*}[t]
\centering
\caption{Summary of DM-enabled DRL for Resource Allocation, Security, and Routing}
\label{Tab_Sec_IV}
\renewcommand{\arraystretch}{1.2}

\begin{tabular}{|l|l|l|l|l|}
\hline
\textbf{Category} & \textbf{Ref.} & \textbf{Objective} & \textbf{Proposed Method} & \textbf{Role of Diffusion} \\
\hline

\multirow{8}{*}{\makecell[l]{Power Allocation}}
& \cite{wu2025drl} & Spectral Efficiency Maximization & DM-enabled Offline DRL & Expert Trajectory Generator \\ 
\cline{2-5}
& \cite{zhang2025improve} & Sum-Rate Maximization & DM-enabled DDPG & State-Action-Reward Explorer \\ 
\cline{2-5}
& \cite{zhang2025enhanced} & Secrecy Rate Maximization & DM-enabled Actor-Critic & Policy Generator \\ 
\cline{2-5}
& \cite{wang2025uplink} & Sum-Rate Maximization & DM-enabled Hybrid PPO & Action Distribution Generator \\ 
\cline{2-5}
& \cite{wang2024generative} & Sum-Rate Maximization & DM-enabled PPO & Continuous Policy Generator \\ 
\cline{2-5}
& \cite{xie2025multi} & Rate and Echo Maximization & DM-enabled DDPG & Action Policy Generator \\ 
\cline{2-5}
& \cite{li2024dfrl} & Energy Consumption Minimization & DM-enabled RL & Discrete Policy Generator \\ 
\hline

\multirow{8}{*}{\makecell[l]{Spectrum Allocation}}
& \cite{khoramnejad2025carrier} & Load Balancing Optimization & DM-enabled Actor-Critic & Carrier Selection Generator \\ 
\cline{2-5}
& \cite{xiong2025diffusion} & PRB Usage Minimization & DM-enabled SAC & Multi-Modal Policy Generator \\ 
\cline{2-5}
& \cite{shi2025diffusion} & Delay and Throughput Optimization & DM-enhanced DQN & Trajectory Planner \\ 
\cline{2-5}
& \cite{darabi2024diffusion} & Reliability Maximization & DDPM-based Optimization & Optimization Solution Generator \\ 
\cline{2-5}
& \cite{nouri2025diffusion} & Joint PRB and Power Allocation & DM-enabled Q-Learning & Conditional Action Generator \\ 
\cline{2-5}
& \cite{ning2025diffusion} & Channel and Power Allocation & DM-enabled Hierarchical DRL & Hierarchical Policy Generator \\ 
\cline{2-5}
& \cite{meng2025multi} & Delay and Packet Loss Minimization & DM-enabled Multi-Agent RL & Conditional Trajectory Generator \\ 
\hline

\multirow{5}{*}{\makecell[l]{Security\\ and Privacy}}
& \cite{zhang2024multi} & Secure UAV Transmission & DM-enabled TD3 & Action Generator \\ 
\cline{2-5}
& \cite{liang2024uav} & Secure UAV AoI Minimization & DM-enabled TD3 & Robust Actor Policy \\ 
\cline{2-5}
& \cite{you2024generative} & VR-AIGC Offloading & DM-enabled PPO & Privacy-Preserving State Transform \\ 
\cline{2-5}
& \cite{kang2025hybrid} & Secure Vehicular Twin Migration & DM-enabled DRL & Robust Migration Policy \\ 
\cline{2-5}
& \cite{chen2025trust} & Trust-Aware Vehicular Consensus & DM-enabled SAC & Consensus Action Generator \\
\hline

\multirow{11}{*}{\makecell[l]{Robotics Control,\\ Vehicle Routing,\\ and UAV Planning}}
& \cite{li2023crossway} & Robot Manipulation & DM-enabled Behavioral Cloning  & Action-Sequence Policy \\ 
\cline{2-5}
& \cite{wang2025integrating} & Offline Robot Manipulation & DM-enabled Q-Guided Offline RL & Vision-Language Action Policy \\ 
\cline{2-5}
& \cite{dong2024diffuserlite} & Real-Time Offline Robot Planning & DM-enabled Offline RL & Trajectory Planner \\ 
\cline{2-5}
& \cite{he2023diffusion} & Robotic Manipulation & DM-enabled Multi-Task Offline RL & Planner and Data Synthesizer \\
\cline{2-5}
& \cite{ho2024team} & Robotic Arm Control & DM-enabled Clustered Meta-RL & Task Representation and Planning \\
\cline{2-5}
& \cite{qiao2025combined} & Soft-Window Vehicle Routing & DM-enabled Actor-Value RL & Feasible Route Generator \\ 
\cline{2-5}
& \cite{hu2025toward} & Multi-Task Autonomous Driving & DM-enabled SAC & Driving Policy Network \\
\cline{2-5}
& \cite{emami2025diffusion} & UAV Communication Control & DM-enabled DT-RL & Generative Decision Tool \\ 
\cline{2-5}
& \cite{yu2025multi} & Multi-UAV Trajectory Control & DM-enabled SAC & Action-Policy Generator \\
\cline{2-5}
& \cite{zhao2024energy} & UAV--UGV Cooperative Sensing &  DM-enabled Multi-Agent DRL & High-Level Navigation Policy \\
\hline

\end{tabular}
\end{table*}

\subsection{Power Allocation}

Traditional DRL has inherent limitations in wireless network power allocation. These include heavy reliance on online deployment, poor generalization to dynamic channel state distributions, high computational costs from periodic retraining, inefficiency in handling hybrid action spaces, and difficulty meeting multi-objective trade-offs or scenario-specific QoS requirements. To fill these gaps, DM-enhanced RL frameworks have emerged. They leverage DMs’ denoising capabilities, diverse sample generation, and distribution modeling to boost DRL’s adaptability. This is critical for real-time, reliable wireless systems, as suboptimal power decisions can disrupt services or waste resources.

To tackle QoS variability and state information exposure in general wireless networks (e.g., 6G, Wi-Fi), \cite{wu2025drl} proposes a wireless network intent-guided trajectory generation model based on DM. This framework enables real-time fine-tuning to align with target network objectives, reduces dependence on online training, and enhances spectral efficiency stability. It achieves this by customizing power allocation and resource management to scenario-specific QoS demands. Compared with traditional DRL baselines, the DM-enabled DRL achieves up to 9.3\% improvement in spectral efficiency. To address DRL’s core bottlenecks in dynamic resource allocation, \cite{zhang2025improve} introduces DM-enabled DRL. These bottlenecks include high-dimensional state spaces, suboptimal exploration-exploitation balance, and inadequate reward design. The proposed DM-enabled DRL scheme reduces retraining costs by over 58\% and accelerates convergence through reward space exploration and diverse state sample synthesis.

Multiuser IoT systems face challenges from channel spatial correlation and imperfect CSI, which harm beamforming-based secure power allocation. To solve this, \cite{zhang2025enhanced} develops an IRS-assisted secure beamforming scheme integrated with a DM-enabled actor-critic algorithm. This framework jointly optimizes base station precoding matrices and IRS phase shifts. It uses DM’s denoising to recover optimal beamforming solutions from channel noise, and ultimately outperforms artificial noise and conventional beamforming schemes. Turning to LEO terrestrial-satellite networks with uplink rate splitting multiple access (RSMA), DRL struggles with hybrid discrete-continuous actions and irregular environments. Thus, \cite{wang2025uplink} formulates a parameterized MDP to convert discrete actions into continuous vectors. It then embeds DMs into hybrid PPO to improve sample quality and policy stability, and delivers larger rewards across varying terminal numbers and power budgets. The proposed DM-enabled hybrid PPO scheme improves the sum rate by 14.52\% compared with the non-DM baseline.

In vehicular networks, minimizing AoI and transmit power via non-orthogonal multi-modal dissemination forms a mixed-integer nonlinear programming problem. To address this, \cite{al2024non} decomposes the multi-objective task using a weighted-sum approach. A hybrid DQN-DDPG model optimizes decoding order via DQN and continuous power allocation via DDPG. Meanwhile, a two-stage meta-multi-objective RL solution reduces retraining overhead, which enables high-quality Pareto front estimation without full model retraining. Complementing LEO-RSMA research, \cite{wang2024generative} models long-run sum data rate maximization as a MDP and proposes a DM-enabled PPO framework. Notably, this framework uses DM’s denoising network as PPO’s actor. This simplifies hyperparameter tuning, boosts sample efficiency, and enhances joint power control and beamforming performance. The proposed DM-enabled PPO scheme improves the sum data rate by 25.87\% compared with the non-DM RSMA baseline.

For ISAC systems that need balanced communication, sensing, and energy efficiency, \cite{xie2025multi} proposes a DM-enabled DDPG method. Integrated with an IRS-UAV to improve channel quality, the DM-enabled DDPG embeds DM into DDPG’s actor network. It uses noise perturbation for exploration and recent prioritized experience replay for training efficiency, and optimizes power-related parameters (e.g., beamforming and UAV propulsion energy). This achieves superior performance across communication rate, target echo rate, and energy metrics. Finally, to address 5G/6G base station energy consumption amid rising mobile traffic, \cite{li2024dfrl} develops a discrete action diffusion RL algorithm. It maps DM’s forward (noise addition) process to RL training and reverse (denoising) process to decision-making. A customized reward function prioritizes energy savings while maintaining QoS. Collectively, these DM-RL frameworks unify solutions to key challenges in wireless power allocation, span diverse scenarios, and lay the groundwork for future intelligent networks. The proposed DM-enabled RL scheme saves 20\%--42.5\% energy compared with popular DRL algorithms while maintaining an acceptable QoS level.

DM-enhanced RL effectively mitigates three critical flaws of traditional DRL in power allocation (poor generalization to dynamic channel conditions, inefficient handling of hybrid discrete-continuous actions, and high retraining overhead) by leveraging DMs’ denoising capability, diverse sample generation, and distribution modeling, making it well-suited for dynamic, resource-constrained systems like LEO networks (with time-varying user access) and IoT systems (with imperfect CSI); however, it relies on idealized assumptions (e.g., sufficient initial training data, stable noise characteristics) and may require frequent parameter adjustments in highly volatile environments (e.g., dense urban vehicular networks), eroding partial efficiency gains.

\subsection{Spectrum Allocation}

Existing DRL-based spectrum allocation and radio resource management approaches have demonstrated strong sequential decision-making capability, yet they remain fundamentally constrained by uni-modal policy representations, unstable exploration in high-dimensional action spaces, and limited adaptability to dynamic wireless environments. These limitations become particularly evident in modern wireless systems, where near-optimal resource configurations are inherently multi-modal, discontinuous, and highly coupled across spectrum, power, routing, and traffic dimensions. Consequently, conventional DRL methods often suffer from unstable convergence, poor generalization, non-stationary training behaviors, and suboptimal resource coordination in real-time systems. To overcome these challenges, recent studies have increasingly integrated DMs into DRL frameworks, leveraging their powerful generative capability to capture complex action distributions and enable more robust decision-making under uncertainty.

\begin{figure}[t]
    \centering
    \includegraphics[width=\linewidth]{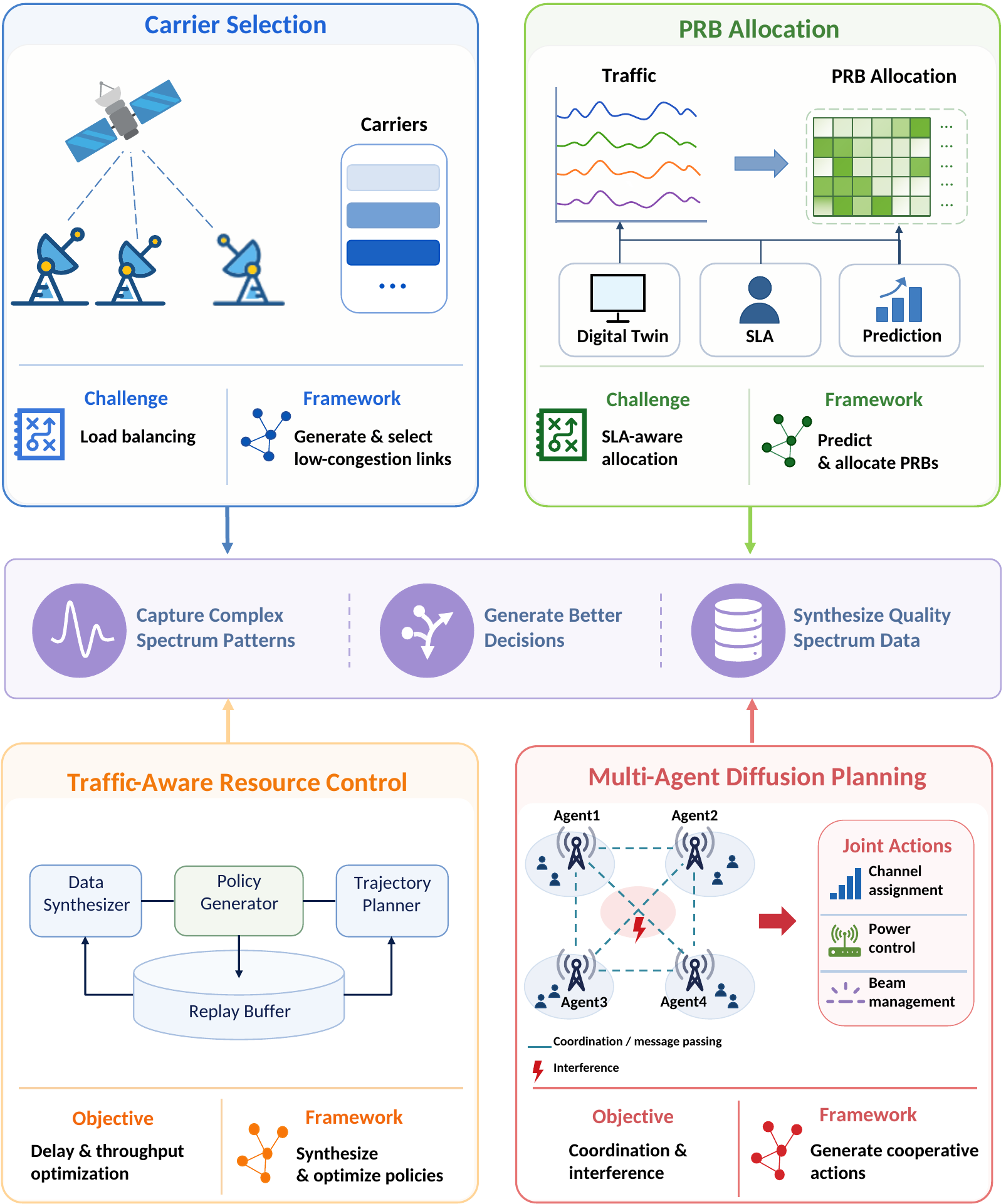}
    \caption{Applications of diffusion models in spectrum allocation and wireless resource control.}
    \label{fig:spectrum_dm}
\end{figure}

As shown in Fig.~\ref{fig:spectrum_dm}, DMs have been applied to spectrum allocation in multiple forms, including carrier selection, PRB allocation, traffic-aware resource control, and multi-agent diffusion planning. By generating low-congestion links, predicting SLA-aware PRB demands, synthesizing traffic-resource samples, and coordinating joint actions among agents, DMs can better capture complex spectrum patterns and dynamic interference, thereby improving resource utilization, decision robustness, and adaptability.

One important research direction focuses on exploiting diffusion models to improve action generation and policy optimization for spectrum and resource allocation. For instance, the authors in \cite{khoramnejad2025carrier} investigate carrier aggregation and backhauling optimization in non-terrestrial network. By integrating DDPM with an actor-critic DRL framework and multi-armed bandit mechanisms, the proposed approach performs iterative denoising to progressively refine carrier selection strategies. The generated policies effectively capture complex resource allocation dependencies. Compared with DDQN baselines, the framework significantly reduces component carrier load and converges substantially faster, highlighting the advantage of diffusion-guided exploration in highly dynamic spectrum environments.

Similarly, diffusion-enhanced DRL has been employed to address the multi-modal nature of physical resource block (PRB) allocation in network slicing systems. The work in \cite{xiong2025diffusion} proposes a conventional generative diffusion-enhanced soft actor critic (CGDSAC) framework to model latent traffic distributions and generate effective PRB allocation strategies. Unlike traditional SAC or PPO methods that rely on unimodal policy distributions, the diffusion process enables the policy network to represent multiple feasible allocation patterns under varying traffic conditions. Furthermore, the integration of a digital twin network, behavior cloning, and KPI prediction mechanisms improves training safety and accelerates convergence during early-stage exploration. As a result, the framework achieves substantially lower SLA violation rates and reduced PRB consumption compared with several conventional DRL baselines.

Beyond direct action generation, diffusion models have also been utilized to enhance DRL through synthetic data generation and trajectory planning. The work in \cite{shi2025diffusion} integrates diffusion models into a DQN-based framework through three dedicated modules, namely a data synthesizer, a policy generator, and a trajectory planner. The data synthesizer enriches replay buffers with high-quality multi-modal samples, thereby alleviating the limited exploration capability of conventional DRL. Meanwhile, the conditional diffusion-based policy generator captures intricate state-action dependencies, enabling more expressive policy representations in complex traffic control scenarios. Experimental results demonstrate significant gains in delay, throughput, and packet loss performance compared with traditional DQN variants and Open Shortest Path First (OSPF) routing strategies.

Another line of research investigates replacing reward-driven DRL optimization entirely with diffusion-based generative learning. The authors in \cite{darabi2024diffusion} propose a DDPM-based framework for joint blocklength selection, sampling period optimization, and packet error probability control in ultra-reliable wireless networked control systems (WNCSs). Instead of relying on iterative reward exploration, the DDPM directly learns the distribution of optimization-theoretic solutions from channel state information (CSI). This design avoids the instability and infeasibility issues commonly encountered in DRL approximation processes while substantially reducing constraint violations. Moreover, the diffusion framework achieves significantly higher reliability and improved GPU utilization efficiency compared with state-of-the-art DRL solutions, illustrating the potential of DMs as direct policy generators for wireless optimization problems.

Diffusion-guided Q-learning has also emerged as a promising solution for addressing the uni-modal action limitations of traditional value-based DRL methods. The work in \cite{nouri2025diffusion} proposes a DM-enabled Q-learning framework for joint power and PRB allocation, where resource allocation actions are iteratively generated through conditional denoising. In each denoising step, gradients derived from a learned Q-function guide the refinement process toward high-quality actions. By modeling the resource allocation problem as a conditional diffusion process, the framework effectively captures multiple near-optimal action configurations and demonstrates superior approximation accuracy relative to exhaustive search solutions. The results reveal notable improvements in MAE, cosine similarity, and binary association accuracy, validating the effectiveness of diffusion-guided policy generation in high-dimensional wireless resource allocation tasks.

The capability of diffusion models to capture hierarchical and large-scale resource dependencies has further motivated their adoption in multi-level spectrum management architectures. The work in \cite{ning2025diffusion} proposes a hierarchical DRL framework that jointly optimizes communication quality and collaborative efficiency in construction equipment networks. Specifically, a DM-enabled SAC algorithm is utilized at the high-level controller to capture global spectrum allocation patterns and maintain stable long-term reward trajectories, while QMIX is employed for intra-cluster resource coordination. By combining hierarchical decision-making with diffusion-based global policy generation, the framework effectively mitigates non-stationarity and combinatorial action explosion problems that typically degrade conventional single-layer DRL approaches.

In addition to hierarchical optimization, diffusion-enhanced multi-agent DRL has demonstrated strong scalability advantages in dense wireless environments. The authors in \cite{meng2025multi} introduce a multi-agent conditional diffusion model planner (MACDMP), which combines diffusion models with model-based reinforcement learning for conditional trajectory generation. Rather than relying on global information exchange among all agents, the framework leverages local observations and mean-field guidance to model neighborhood interactions during the denoising process. This design significantly improves scalability and coordination efficiency under dynamic RF conditions while reducing communication overhead. Experimental results show lower delay, reduced packet loss, faster convergence, and higher peak rewards under both ideal and constrained RF scenarios, demonstrating the robustness of diffusion-guided trajectory planning in large-scale wireless systems.

Overall, these studies collectively demonstrate that diffusion models fundamentally reshape the design paradigm of DRL-enabled wireless resource management. By leveraging iterative denoising mechanisms, conditional generative modeling, and multi-modal distribution learning, DM-enabled DRL frameworks overcome several intrinsic limitations of conventional DRL, including inefficient exploration, unstable convergence, and limited policy expressiveness. More importantly, diffusion models enable wireless decision-making systems to better capture the highly dynamic, discontinuous, and combinatorial nature of spectrum allocation and resource optimization problems. As a result, DM-enabled DRL has emerged as a highly promising research direction for next-generation wireless networks, providing enhanced robustness, scalability, and decision quality across diverse spectrum management scenarios.


\subsection{Security and Privacy}

Recent studies have integrated DMs into RL to address security and privacy challenges. 
In these methods, DMs serve as generative and transformation components in the decision-making process, enabling security- and privacy-aware resource allocation and control. 
DM-enabled DRL methods have been applied to mitigate adversarial attacks, privacy leakage, and trust uncertainty.

In UAV-assisted wireless systems, DM-enabled DRL has been explored to enhance physical-layer security against eavesdropping and jamming. 
In the presence of mobile eavesdroppers, \cite{zhang2024multi} investigates secure long-range transmission in UAV swarms via collaborative beamforming. 
A generative DM-enhanced TD3 framework is employed to generate control actions for UAV positioning and beamforming. 
Compared with conventional UAV deployment strategies, the DM-enhanced TD3 achieves higher secrecy rates with improved energy efficiency. 
Under active jamming attacks in UAV-enabled IoT systems, \cite{liang2024uav} studies the joint optimization of information freshness and energy efficiency. 
A UAV transfers energy to IoT devices and collects sensing data over jammed wireless channels. 
To minimize secure AoI and UAV energy consumption, a DM–enhanced TD3 framework is proposed, in which a generative DM is integrated into the actor network to improve the robustness of action generation under adversarial conditions. 
The proposed DM–enhanced TD3 framework achieves lower secure AoI and improved energy efficiency.

In VR-based AIGC services, DM-enabled DRL has been applied to mitigate privacy leakage. 
To this end, \cite{you2024generative} studies privacy-preserving computation offloading for VR-AIGC under centralized PPO control. 
A generative DM is introduced to perturb PPO states and transform them into privacy-preserving representations through a denoising process, enabling effective decision making without directly exposing the original state information. 
To evaluate the impact of this state transformation on learning performance, inverse RL is employed to assess the consistency between rewards derived from the original and transformed states. 
The proposed Diffusion-assisted approach preserves VR user privacy with limited performance degradation.

In vehicular networks, DM-enabled DRL has been employed to support security- and trust-aware decision making. 
For vehicular metaverse scenarios, \cite{kang2025hybrid} studies secure vehicle twin migration under edge-side attacks, including distributed denial-of-service and co-residency threats. 
A hybrid generative diffusion–enabled DRL scheme is employed to optimize edge server selection and pre-migration decisions. 
Through iterative denoising, the scheme generates robust migration policies, leading to lower migration latency and the selection of higher-reputation edge servers. 
For vehicle platooning systems, \cite{chen2025trust} studies trust-aware blockchain consensus optimization under heterogeneous trust conditions. 
A role-adaptive trust model is constructed to evaluate vehicle reliability, and the consensus problem is addressed using a Diffusion-enhanced SAC framework.
By embedding a DM into the SAC actor to generate consensus actions via iterative denoising, the proposed scheme reduces consensus latency, increases throughput, and improves primary-node reliability.

\subsection{Robotics Control, Vehicle Routing and UAV Planning}

DM has recently emerged as a promising tool for sequential decision making in mobile systems.
In robotics, autonomous driving, and UAV-enabled systems, DMs are increasingly used as generative decision modules, enabling better control performance, stronger generalization, and more efficient adaptation in dynamic environments.

In \cite{wolf2025diffusion} and \cite{bai5237963diffusion}, DMs are reviewed as an emerging tool for RL in robotics, where they are mainly used as generative policies or planners for robot control and sequential decision making.
Following these survey discussions, \cite{li2023crossway} provides a concrete example of DMs as action-sequence policies for robot manipulation.
A DM-enabled action-sequence policy is adopted in behavioral cloning, where future actions are generated by iterative denoising.
The proposed method improves the task success rate by 15.7\% over the baselines and also performs better on real-world tasks.
Building on this action-sequence policy perspective, \cite{wang2025integrating} further studies offline robot manipulation learning from language-driven datasets containing both success and failure trajectories.
A vision-language-conditioned diffusion policy is developed to generate temporally consistent action sequences from visual observations and language instructions, and is further refined with Q-value guidance in offline RL.
The proposed method outperforms existing baselines in success rate and achieves comparable performance with 20\%–30\% less success data.
Moving beyond direct action-sequence policy learning, \cite{Kapelyukh2022DALLEBotIW} studies zero-shot robotic object rearrangement in tabletop scenes.
A web-scale DM is introduced as an image-generation module to produce a human-like goal arrangement from object descriptions, which is then converted into target poses for pick-and-place rearrangement.
Taking a step further from goal generation to explicit planning, \cite{dong2024diffuserlite} studies real-time diffusion planning for offline RL across locomotion, robot manipulation, navigation, and stock-trading tasks.
A DM-enabled trajectory planner is developed in a progressive refinement process, where coarse-to-fine state sequences are generated and then converted into actions with an inverse dynamics model.
The proposed method improves decision frequency by over 112 times while maintaining state-of-the-art performance.
Beyond single-task settings, \cite{he2023diffusion} studies multi-task offline RL for robotic manipulation and navigation tasks.
The DM is used as both a planner and a data synthesizer, generating action sequences for multi-task decision making and high-fidelity transitions for offline augmentation, while SAC and TD3 with behavior cloning are used for data collection and policy evaluation, respectively.
The proposed method achieves up to 180\% improvement in offline policy performance on unseen tasks.
Further emphasizing task adaptation, \cite{ho2024team} studies multi-task meta-RL for robotic arm control.
A DM-enabled multi-task representation and planning module is introduced, where task-specific representations are generated through reverse denoising and then combined with performance-driven task clustering and fast adaptive multi-task optimization.
The proposed framework improves the task success rate to 93.2\% while effectively enhancing training efficiency.

\begin{figure}[htbp]
    \centering
    \includegraphics[width=\linewidth]{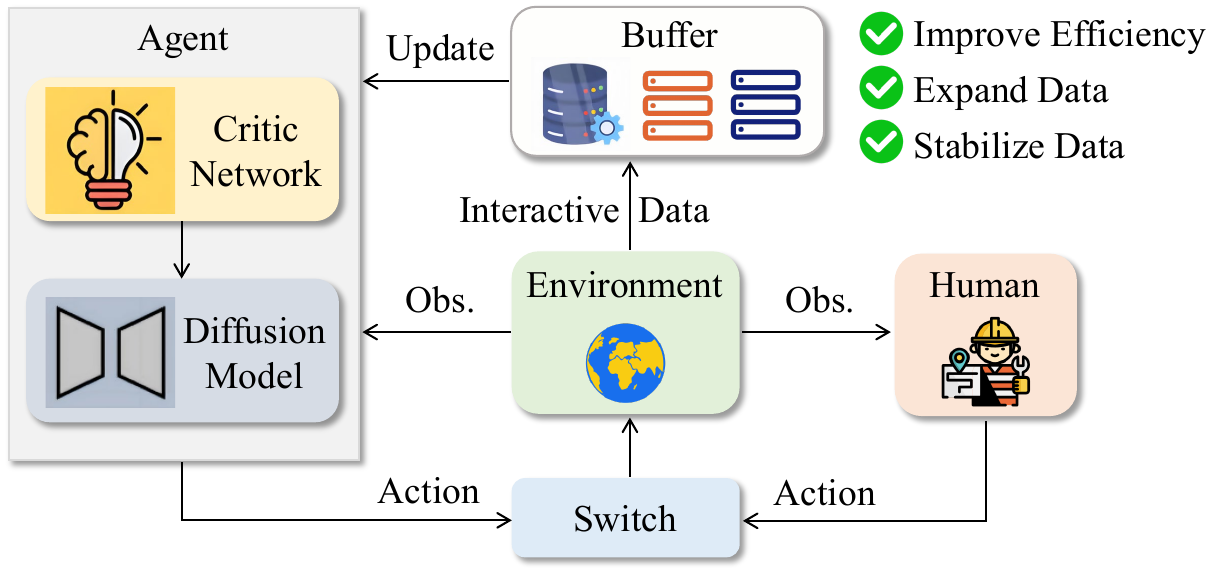}
    \caption{The diagram of the proposed human-in-the-loop RL with diffusion policy in \cite{hu2025toward}.}
    \label{Fig_human-in-the-loop_RL}
\end{figure}

For vehicle-related tasks, recent studies have explored DM-enhanced RL for both route optimization and autonomous navigation.
In \cite{qiao2025combined}, the authors study the vehicle routing problem with multiple soft time windows, where vehicles must serve customers under capacity, distance, and soft time-window constraints.
A DM-enabled path generator is used to produce feasible routing paths through iterative denoising, and the actor-value-style RL module further refines these routes.
The proposed method achieves up to 9\% lower cost on large-scale instances.
Moving from route optimization to embodied navigation control, \cite{hu2025toward} investigates multi-task autonomous navigation for autonomous ground vehicles in dynamic environments, including urban streets, unsignalized intersections, and pedestrian-dense areas.
A DM-enabled policy network is integrated into an online SAC framework, as shown in Fig. \ref{Fig_human-in-the-loop_RL}, where actions are generated through reverse denoising and the policy is trained with behavior cloning, Q-value guidance, human-in-the-loop feedback, and adversarial learning.
The proposed method achieves strong real-world performance, with improved efficiency, stability, and cross-task generalization over existing baselines.

\begin{figure}[htbp]
    \centering
    \includegraphics[width=0.8\linewidth]{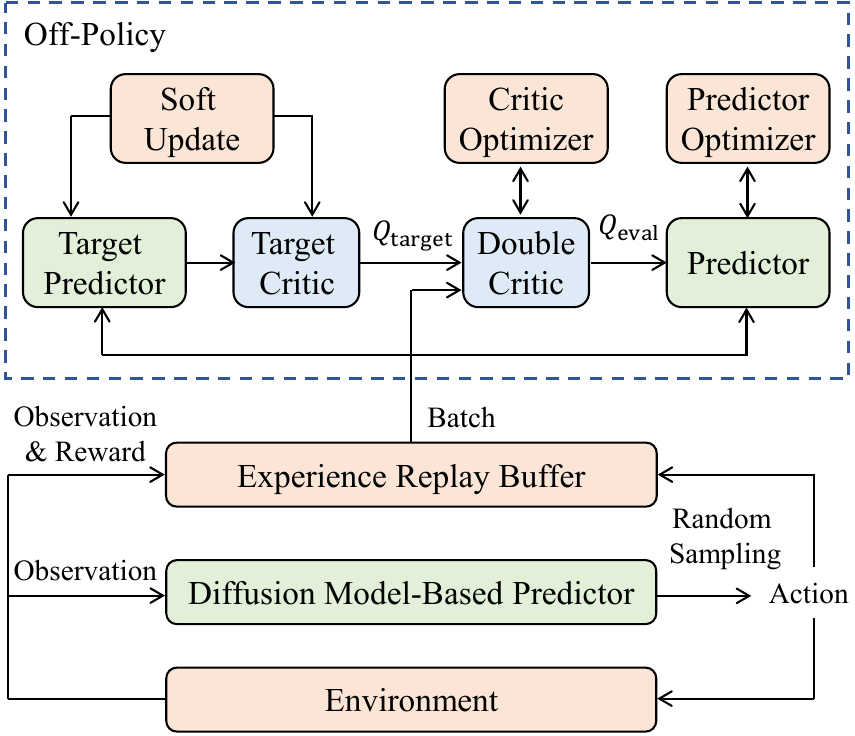}
    \caption{The diagram of the proposed DM-enabled DRL scheme for multi-UAV trajectory generation in \cite{yu2025multi}.}
    \label{Fig_MUTG_algorithm}
\end{figure}

For UAV-related applications, DM-enabled DRL has been explored for both UAV communication and control design and cooperative aerial sensing tasks.
In \cite{emami2025diffusion}, the authors discuss how DMs can enhance UAV communications when integrated with RL and digital twin techniques, highlighting their value as generative tools for improving policy learning, simulated training, and adaptability in complex communication scenarios.
Moving from this general perspective to a concrete UAV control task, \cite{yu2025multi} studies multi-UAV trajectory generation for fresh data collection from mobile users, with the goal of reducing AoI while improving user coverage and the amount of collected data.
A DM-enabled action-policy generator is integrated into an SAC framework, as shown in Fig. \ref{Fig_MUTG_algorithm}, where a DM-enabled predictor produces UAV movement actions.
The proposed method outperforms benchmark schemes, achieving up to 13\% improvement in AoI.
Extending UAV control to heterogeneous ground-air coordination, \cite{zhao2024energy} studies post-disaster ground-air-space vehicular crowdsensing, where UGVs transport and charge UAVs for cooperative data collection from points of interest.
A DM-enabled high-level goal-conditioned policy is adopted for UGV navigation in a hierarchical multi-agent DRL framework, where a discrete diffusion policy is optimized with multi-agent SAC, and the low-level UAV sensing policy is learned with independent PPO.
The proposed method achieves a nearly doubled UAV–UGV cooperation factor and up to 89.7\% data collection ratio.
Beyond vehicle routing and UAV planning, \cite{zhang2024two} extends DM-enabled TD3 to continuous operational control in multi-energy microgrids, highlighting the broader applicability of DM-enabled policies across complex sequential decision-making scenarios.

\subsection{Lessons Learned}

This section summarizes core insights from diffusion model-enabled deep reinforcement learning across resource allocation, security and routing.  For power and spectrum allocation, its ability to model complex multi-modal and hybrid discrete-continuous action spaces breaks the constraints of standard Gaussian policies, delivering consistent performance gains in beamforming design, satellite resource coordination, and large-scale multi-agent scheduling. Its generative capability further enhances exploration efficiency, reduces retraining overhead in dynamic wireless environments, and improves system scalability. For security and privacy, DMs mainly strengthen RL by generating more robust or privacy-preserving decision variables, making them particularly useful when the control policy must remain effective under adversarial interference, privacy constraints, or trust uncertainty. For robotics control, vehicle routing and UAV planning, DMs are most effective when used as expressive action or planning priors, since they improve sequential decision making by generating structured trajectories, goals, or policies that enhance control quality, generalization, and adaptation in complex dynamic tasks.

\section{Conclusions and Future Works} \label{sec:conclusions}

This paper has presented a comprehensive survey of diffusion model-enabled deep 
reinforcement learning (DM-enabled DRL) for resource management in wireless networks. 
We first reviewed the theoretical foundations of DM-enabled DRL, covering key algorithms 
including DM-enabled QL, DM-enabled SAC, DM-enabled TD3, DM-enabled DDPG, DM-enabled PPO, 
and their multi-agent variants, as well as complementary roles such as data augmentation 
and reward shaping. We then systematically surveyed applications of DM-enabled DRL for several emerging issues in wireless networks that consist of computation 
offloading in applications, i.e., MEC, UAV-assisted, vehicular, and AIGC-driven systems, wireless resource 
allocation, physical-layer security, and robotics and UAV planning. It can be observed that DM-enabled DRL consistently demonstrates three compelling advantages over 
conventional DRL methods. First, the iterative denoising process enables expressive 
multi-modal policy representations that capture the complex, discontinuous decision 
boundaries inherent in wireless resource management. Second, the stochastic nature 
of the diffusion process naturally promotes exploration, alleviating the exploitation 
bias and local optima susceptibility of standard Gaussian actors. Third, the 
conditioning flexibility of diffusion models allows seamless integration with 
offline datasets, reward signals, and environmental constraints. These advantages make DM-enabled DRL 
broadly adaptable across heterogeneous network scenarios. Even those, there are still 
open challenges in inference latency, theoretical guarantees, safety 
certification, and real-world deployability, which motivate the future research 
directions outlined below.
\begin{itemize}

\item \textit{Quantization-Aware Diffusion Policies for Resource-Constrained Wireless Networks:}
Deploying DM-enabled DRL on resource-constrained devices such as UAV platforms and MEC 
servers remains challenging due to the significant memory and computational overhead 
of the iterative denoising process. Post-training quantization techniques, such as 
Q-Diffusion \cite{li2023q,shang2023post,so2023temporal,he2023ptqd}, offer a promising path to alleviating this burden 
by reducing the bit-width of denoising network parameters without full model retraining. 
However, unlike image generation, quantized DM-enabled DRL policies need to maintain sufficient 
numerical precision to preserve multi-modal expressiveness and avoid reward degradation 
in dynamic wireless environments. Developing quantization-aware diffusion policy 
frameworks tailored to the latency and reliability requirements of next-generation 
wireless networks therefore represents an important open direction.

\item \textit{Lightweight and Real-Time Diffusion Inference:}
A recurring concern across DM-enabled DRL approaches surveyed in this paper is that the multi-step reverse denoising process introduces non-trivial per-decision latency, limiting applicability in hard real-time scenarios such as MEC offloading and UAV trajectory control. This is also specified in \cite{liu2026lyapunov, zhang2026improve, chen2026towards}. Accelerated sampling techniques including consistency models, flow matching, and progressive distillation, as well as more scalable pipeline configuration strategies \cite{sheng2026adapting}, offer promising paths to retaining the expressive power of DM-enabled policies while drastically reducing computational overhead in time-critical wireless resource management.

\item \textit{Safety-Constrained DM-enabled DRL for Critical Infrastructure:}
Most DM-enabled DRL methods surveyed optimize unconstrained objectives and cannot provide hard guarantees on critical constraints such as maximum power budgets, latency deadlines, or collision avoidance, which are indispensable for controllers in industrial IoT and public-safety networks \cite{chen2026towards}. Integrating constrained MDP frameworks or Lyapunov-based stability mechanisms with diffusion policies, as partially explored in \cite{liu2026lyapunov}, represents a high-priority challenge for the safe and certified deployment of DM-enabled DRL in critical infrastructure.

\item \textit{Multi-Agent DM-enabled DRL with Scalable Coordination:}
While mostly works in this survey adopt single-agent DRL formulations, practical wireless networks involve large populations of coupled decision-makers including multi-UAV deployments and dense edge computing environments. Extending DM-enabled DRL to fully distributed multi-agent settings \cite{meng2026networld}, where each agent maintains a local diffusion policy while exchanging compact coordination signals, is a natural next step whose key challenge lies in designing mechanisms that are both communication-efficient and expressive enough to capture inter-agent dependencies.

\item \textit{Risk-Sensitive and Freshness-Aware DM-enabled DRL:}
Emerging wireless applications covered in this survey, including industrial IoT and real-time status update systems, impose strict requirements on the tail behavior of key metrics such as latency and information staleness. Recent works, e.g., \cite{pan2026diffusion}, demonstrate that combining DM-enabled actors with distributional critics enables principled optimization under coherent risk measures such as conditional value-at-risk (CVaR), which open a promising direction for reliability-critical network control where tail-risk guarantees are as important as mean performance.


\item \textit{Federated and Privacy-Preserving DM-enabled DRL:}
Current DM-enabled DRL frameworks surveyed across MEC, vehicular, and AIGC offloading scenarios \cite{liu2026lyapunov, he2026diffusion, wang2026diffusion} generally assume a centralized training environment with full state observability, which is incompatible with privacy and regulatory constraints in practical deployments. Jointly designing DM-enabled policy learning under federated and differential privacy guarantees \cite{chen2026towards}, while preserving the sample efficiency advantages that diffusion models provide over conventional DRL baselines, represents both a technically demanding and practically important open direction.

\item \textit{Foundation Model and Digital Twin-Assisted DM-enabled DRL:}
A persistent limitation of current DM-enabled DRL systems is their scenario-specificity, where policies trained for one network topology or traffic pattern generalize poorly to new deployments. Digital twins provide a high-fidelity simulation sandbox for safe offline pre-training and stress testing \cite{meng2026networld}, while large language models offer semantic generalization by conditioning the diffusion actor's denoising process on natural language operator intents or multi-modal network telemetry \cite{deng2026snr, huang2026diffusion}. Exploring this direction is compelling for enabling zero-shot adaptation across the heterogeneous AI service network scenarios covered in this survey.

\end{itemize}

\bibliographystyle{IEEEtran}
\bibliography{DM_REF_SHORT}{}


\end{document}